\documentclass[sigconf]{acmart}
\usepackage{cleveref}
\usepackage{listings}
\usepackage[font=small,labelfont=bf]{caption}
\usepackage{enumitem}
\usepackage{soul}
\usepackage{multirow}

\usepackage{def}

\AtBeginDocument{%
  }

\copyrightyear{2025}
\acmYear{2025}
\setcopyright{cc}
\setcctype{by}
\acmConference[MICRO '25]{58th IEEE/ACM International Symposium on Microarchitecture}{October 18--22, 2025}{Seoul, Republic of Korea}
\acmBooktitle{58th IEEE/ACM International Symposium on Microarchitecture (MICRO '25), October 18--22, 2025, Seoul, Republic of Korea}
\acmDOI{10.1145/3725843.3756038}
\acmISBN{979-8-4007-1573-0/2025/10}

\begin{document}



\twocolumn
\setcounter{page}{1}

\title{\pname{}: Enabling Power Gating in Neural Processing Units}


\author{Yuqi Xue}
\email{yuqixue2@illinois.edu}
\affiliation{
  \institution{University of Illinois Urbana-Champaign}
  \city{}
  \state{}
  \country{}
}

\author{Jian Huang}
\email{jianh@illinois.edu}
\affiliation{
  \institution{University of Illinois Urbana-Champaign}
  \city{}
  \state{}
  \country{}
}

\begin{abstract}
The energy efficiency of neural processing units (NPU) plays a critical role in developing sustainable data centers. Our study with different generations of NPU chips reveals that 30\%--72\% of their energy consumption is contributed by static power dissipation, due to the lack of power management support in modern NPU chips. 

In this paper, we present \pname{}, which enables fine-grained power-gating of each hardware component in NPU chips with hardware/software co-design.
Unlike conventional power-gating techniques for generic processors, enabling power-gating in NPUs faces unique challenges due to the fundamental difference in hardware architecture and program execution model.
To address these challenges, we carefully investigate the power-gating opportunities in each component of NPU chips and decide the best-fit power management scheme (i.e., hardware- vs. software-managed power gating).
Specifically, for systolic arrays (SAs) that have deterministic execution patterns, \pname{} enables cycle-level power gating at the granularity of processing elements (PEs) following the inherent dataflow execution in SAs. For inter-chip interconnect (ICI) and HBM controllers that have long idle intervals, \pname{} employs a lightweight hardware-based idle-detection mechanism.
For vector units and SRAM whose idle periods vary significantly depending on workload patterns, \pname{} extends the NPU ISA and allows software (e.g., compilers) to manage the power gating. With implementation on a production-level NPU simulator, we show that \pname{} can reduce the energy consumption of NPU chips by up to 32.8\% (15.5\% on average), with negligible impact on AI workload performance. The hardware implementation of power-gating logic introduces less than 3.3\% overhead in NPU chips.



\end{abstract}



\maketitle

\section{Introduction}
\label{sec:intro}
Hardware accelerators for AI workloads, such as neural processing units (NPUs), are pervasive in modern cloud platforms~\cite{cloudtpu:google,aws_inferentia,tenstorrent}.
For instance, Google has produced seven generations of TPUs (a typical example of NPU chips) for its cloud in the past six years to accommodate the growing AI demands~\cite{tpucloud, tpuv4:isca23,ai_chip_shipments}.
This surge in AI has become the major source of the increasing energy consumption of data centers, leading to a significant increase in data center greenhouse gas emissions over the past five years~\cite{reuters2025,time2025,wsj2024,baselinemag2024,google_energy2024}.

The energy efficiency of NPU chips is playing a critical role in the sustainable development of data centers. Unfortunately, NPU chips today suffer from severe energy inefficiency.
Taking the typical TPU architecture as an example, we conduct a characterization study of its energy efficiency across different generations of chips by running various ML workloads, including large language models (LLMs), recommendation models, and stable diffusion models. We report the static and dynamic energy consumption of core hardware components of the chips in $\S$\ref{sec:design_pg:study}, including systolic arrays (SAs), vector units (VUs), on-chip SRAM, HBM controller \& PHY (physical layer), inter-chip interconnect (ICI) controller \& PHY, and others. 


Our study reveals that despite the energy efficiency improvements of newer NPU generations, there still exists a significant energy waste. Given a typical 60\% utilization (duty cycle) for the NPU chips~\cite{meta:sustainable_ai:mlsys2022} and a datacenter power usage efficiency (PUE) of 1.1~\cite{google_pue}, 
17\%--32\% of energy is wasted due to chip idleness (e.g., while the chip is powered on but not running any jobs).
When the chips are busy, 30\%--72\% of energy comes from static power dissipation, due to the lack of system and architectural support for flexible power management in NPU chips. As most machine learning (ML) workloads are either compute-, memory-, or network-bound, they cannot always fully utilize all components on an NPU chip.
Those underutilized components dissipate a significant amount of static power, as the current NPU design cannot dynamically turn them off or put them into low-power mode (i.e., power gating).

Since static power dissipation will be (relatively) higher as the feature size continues to shrink in the future, supporting flexible power management is highly desirable for a sustainable NPU chip design.
A promising approach is to reduce their static power consumption by power-gating idle hardware components at fine granularity. However, given the following unique challenges and opportunities, enabling power-gating in NPU chips is not straightforward.

First, the NPU architecture has unique components such as large systolic arrays. 
The large components will have a long wake-up delay, rendering the conventional idle-detection or no-access policies (developed for generic CPU/GPU architectures)~\cite{warped_gates:micro13,drowsy_cache:isca02} less effective. 
The new components require new techniques for power gating.

Second, the NPU architecture adopts an in-order pipeline without dynamic scheduling to simplify hardware design. It relies on the compiler to exploit the instruction-level parallelism (ILP) of ML programs. This renders many sophisticated power-gating techniques on CPUs/GPUs infeasible, such as using branch prediction and cache hit/miss as hints to decide when to power on/off the ALUs and caches~\cite{pg_alu:islped04,itap,charstar}, and changing the instruction scheduling policy to create longer idle periods for ALUs and cachelines~\cite{warped_gates:micro13,select_devectorization,gpu_inst_reorder:islped16}.

Third, NPUs are domain-specific accelerators for ML programs, which are known to be deterministic and predictable.
This creates unique opportunities for software to manage the power states of individual components on an NPU chip in an accurate manner. 

We present \pname{}, a hardware-software co-design that enables fine-grained power gating of each component in an NPU chip. To identify the power-gating opportunities in NPU chips, we first analyze the temporal and spatial utilization patterns of each component, based on which we decide its best-fit power management scheme (i.e., hardware- vs. software-managed power gating). 
We list our key insights and contributions as follows.

\noindent
\textul{Hardware-managed power gating} is the most effective approach for the SA, the ICI, and the HBM controllers. The SA can be both temporally and spatially underutilized, and ideally, we wish to power gate at the processing element (PE) granularity. A na\"ive idle-detection mechanism that controls the entire SA is insufficient for exploiting the idleness of the individual PEs. We also cannot simply add an idle-detection state machine to each PE due to the high hardware cost.
It is also impractical for software to control the power mode of each PE at the cycle level.
Our key insight is that the SA has a deterministic systolic dataflow pattern across its PEs.
We can follow the dataflow to propagate the power-on/off signal through the PEs.
This enables cycle-level power gating for each PE with minimal hardware overhead.

The ICI is only active in collective operators (e.g., AllReduce, AllToAll, ReduceScatter, and AllGather), and it can be completely power-gated in other operators.
As an operator is typically long enough (at least a few $\mu$s), it suffices to employ an idle-detection mechanism for power gating the ICI controller \& PHY. Similarly, the HBM controller will observe long idle intervals for compute-intensive operators that favor large tile sizes for improved arithmetic intensity. In such cases, we can power off the DMA engine and switch the HBM controller to low-power auto-refresh mode.



\noindent
\textul{Software-managed power gating} is the most effective solution for the VU and the SRAM.
VU idleness is mostly due to waiting for other components (e.g., SA, HBM, and ICI), and the exact idle periods vary significantly depending on the workload patterns and compiler optimizations (e.g., operator fusion and instruction scheduling). 
Therefore, instead of relying on a hardware-based idle-detection mechanism, we leverage the compiler to decide when to power on/off the VU based on the distance between two VU instructions. This is feasible, as NPUs usually employ a statically scheduled VLIW ISA and rely on the compiler to perform instruction scheduling.

The SRAM capacity utilization varies for different tensor operators as they have different tile size demands. Ideally, we can power gate the unused SRAM capacity during the execution of the entire operator
(i.e., with the gated-Vdd approach~\cite{gated-vdd, gatedvdd:islped00}) and do not need to retain any data in the SRAM cells.
However, this is impossible without software control since the hardware has no hints about the tile size or the data layout in the SRAM.
A possible hardware-only solution, originally designed for the CPU cache, is to apply the idle-detection mechanism to each SRAM segment (4KB in our NPU) and put idle partitions into a low-power mode~\cite{drowsy_cache:isca02}. The low-power mode retains data in the SRAM cells, but it still consumes larger leakage power than a complete power-off.
To maximize the energy savings, we leverage the ML compiler to decide how much of the SRAM capacity should be power-gated for an operator.

\noindent
\textul{ISA extension for flexible power management.}
To support software-managed power gating, we extend the ISA of NPUs to expose the power management functions to the software, such as power gating an execution unit or a portion of SRAM. For each component, we provide new instructions to switch between three power states: on, auto, and off. The auto state is enabled by default, in which the hardware transparently manages power gating. The on/off states override the hardware policies and enable the software (e.g., ML compiler) to customize the power gating policy.


We synthesize the proposed power gating hardware with Cadence Genus and place \& route the design with Cadence Encounter.
The hardware implementation of power-gating logic introduces less than 3.3\% area cost in NPU chips.
We also verified the functionality of \mbox{\pname{}} with FPGA.
We examine the energy/power savings and the performance overhead of \pname{} with a production-level NPU simulator.
\pname{} can reduce the energy consumption of ML workloads by 8.5\%--32.8\% (15.5\% on average) compared to no power gating, while the performance degradation is less than 0.5\%.
At scale, \pname{} reduces the operational carbon emissions of the NPU fleet by 31.1\%--62.9\%.
We list our contributions as follows:
\begin{itemize}[leftmargin=*,itemsep=1ex]
    \item To our best knowledge, we are the first to quantify the power-gating opportunities across all major components in NPU chips.
    \item We enable architectural support for power gating each component on an NPU chip, and identify the best-fit power-gating mechanism (hardware-based or software-based). 
    \item We extend the ISA of NPU chips to enable software support for managing NPU chip power gating at fine granularity based on AI workload characteristics.
    \item We evaluate \pname{} with a production-level NPU simulator, showing that \pname{} can significantly improve the energy efficiency of NPU chips with minimal performance overhead.
    \item We verify the RTL implementation of the power-gating logic on FPGA. We also synthesize and place \& route the design with a 7nm PDK, showing that \pname{} introduces little hardware cost. 

\end{itemize}

\section{Background and Motivation}
\label{sec:background}

\subsection{System Architecture of NPUs}
\label{sec:bkg:npu_arch}

\begin{figure}[t]
    \centering
    \includegraphics[width=0.9\linewidth]{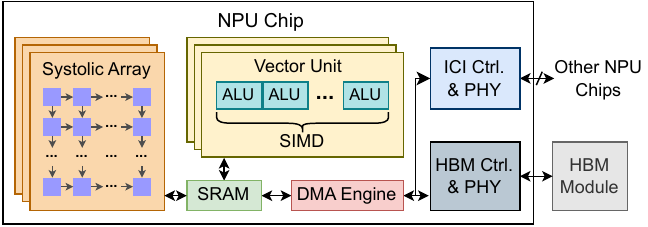}
    \caption{Architecture of an NPU chip.}
    \label{fig:npu_arch}
\end{figure}

Neural processing units (NPUs) are specialized accelerators for machine learning (ML) workloads.
A typical example is Google TPU~\cite{tpuv4:isca23}, the state-of-the-art NPU architecture widely deployed in production.
\Cref{fig:npu_arch} shows the NPU chip architecture.
It consists of systolic arrays (SAs) for matrix multiplications and SIMD vector units (VUs) for generic vector operations.
Each chip has an off-chip high-bandwidth memory (HBM) to store the ML model weights and input/output data, and an on-chip SRAM to exploit data locality and hide HBM access latency.
A direct memory access (DMA) engine performs asynchronous memory copy between the HBM and SRAM.
Multiple NPU chips can be connected via high-speed inter-chip interconnect (ICI) links, which form an NPU pod. A pod is typically arranged as a 2D/3D torus, which is optimized for all-reduce bandwidth~\cite{tpuv4:nsdi24}. The DMA engine performs remote DMA (RDMA) operations to access another chip's HBM or SRAM.



\begin{table}[t]
    \centering
    \caption{Benchmark ML workloads studied in this paper.}
    \vspace{-2ex}
    \footnotesize
    \begin{tabular}{|c|c|c|}
        \hline
        Application & ML Model & Default Config. \\\hline
        \multirow{2}{*}{LLM Training} & \multirow{5}{11em}{\centering Llama3-8B~\cite{llama3}, Llama2-13B~\cite{llama2}, Llama3-70B, Llama3.1-405B} & Batch Size: 32; \\
        & & Seq. Len.: 4096 \\ \cline{1-1} \cline{3-3}
        \multirow{3}{6em}{\centering LLM Inference Prefill/Decode} & & Batch Size: 1; \\
        & & Input Seq. Len.: 4096; \\
        & &  Output Seq. Len.: 512 \\ \hline
        \multirow{3}{*}{Recommendation} & \multirow{3}{11em}{\centering DLRM-S~\cite{DLRM19}, DLRM-M~\cite{tensorflow-model-garden}, DLRM-L~\cite{reddi2019mlperf}} & Batch Size: 1024; \\
        & & Embedding Table Size \\
        & & (S/M/L): 20/45/98 GB \\ \hline
        Stable Diffusion & DiT-XL~\cite{dit}, GLIGEN~\cite{gligen} & Image: 512$\times$512 \\ \hline 
    \end{tabular}
    \label{tab:workloads}
\end{table}

\begin{figure*}[t]
    \centering
    \includegraphics[width=\linewidth]{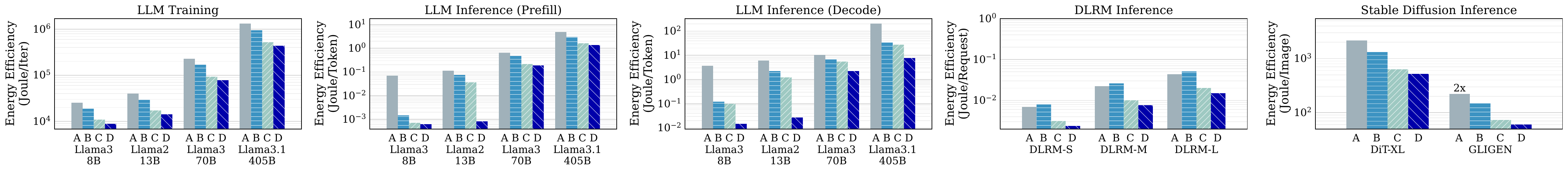}
    \vspace{-4ex}
    \caption{Energy efficiency of ML workloads on different NPU generations. For some NPU generations that cannot satisfy 1$\times$ SLO, we report the energy efficiency for the best relaxed SLO target they can achieve, and the attainable SLOs are labeled on top of the bars (e.g., ``2$\times$'').}
    \label{fig:energy_efficiency}
\end{figure*}

\subsection{Power Gating Basics}

Power gating is a technique to reduce the leakage current flowing through a circuit block when the circuit is not being used.
In general, there are two types of power gating techniques.

The first type completely powers off the circuit block and does not retain any data if the circuit contains flip-flops or SRAM cells. A typical implementation is Gated-Vdd~\cite{gatedvdd:islped00}. It is implemented by adding a header transistor between the supply voltage (Vdd) and the circuit block or a footer transistor between the circuit block and the ground (GND).
This technique is commonly employed for power-gating execution units, which do not need to retain data~\cite{warped_gates:micro13}.

The second type puts the circuit block into sleep (drowsy) mode by applying a reduced Vdd.
The sleep mode consumes more leakage power than a power-off, but it allows the circuit to retain data during sleep mode.
When the data needs to be accessed, the circuit block must first switch back to active mode.
This technique is commonly employed in CPU/GPU caches that need to retain data~\cite{drowsy_cache:isca02}.

\subsection{Power Gating Challenges in NPUs}

Since powering on/off the circuit takes extra latency and consumes dynamic power, the idle period must be larger than the \textit{break-even time} (BET) to avoid negative impact on performance and energy~\cite{warped_gates:micro13}.
Therefore, having long idle periods and accurately detecting idleness are critical to the effectiveness of power gating.
Traditional CPU/GPU architectures tackled these challenges by predicting the idle periods using hints such as branch prediction results and cache hit/miss. Prior works also proposed creating longer idle periods by instruction scheduling~\cite{warped_gates:micro13}, exploiting the unused cache capacity by periodically putting all cachelines into sleep mode and waking up a cacheline upon access~\cite{drowsy_cache:isca02}, and dynamically resizing the cache to power off the unused portion without retaining data~\cite{gatedvdd:islped00}.

As the NPU has a different architecture from CPUs and GPUs, it faces unique challenges for power gating.
First, its SA is much larger than the INT/FLOAT ALUs in CPUs/GPUs.
It takes much longer to power off/on the SA, resulting in a large BET that conventional idle-detection policies cannot easily exploit.
Also, the SA consists of many processing elements (PEs) that can be idle at different cycles, necessitating a more granular power-gating mechanism.
Second, the SRAM is used as a software-managed scratchpad memory rather than a hardware-managed cache.
We cannot easily resize the SRAM transparently without semantic knowledge from the upper-level software.
Third, the NPU core uses an in-order pipeline and a statically scheduled VLIW ISA to simplify hardware design, relying on the compiler to perform instruction scheduling to exploit ILP. 
We cannot directly apply conventional idle period prediction and instruction scheduling techniques in NPUs.

\begin{table}[t]
    \centering
    \caption{NPU specifications used in our study. NPU-A/B/C/D are derived from TPUv2/3/4/5p. NPU-E is a projected NPU generation corresponding to TPUv6p. Parameters with an asterisk (*) are inferred from public data, as they are not officially disclosed for TPUs.
    }
    \vspace{-2ex}
    \footnotesize
    \begin{tabular}{|c|p{3.6em}|p{3.6em}|p{3.45em}|p{3.45em}|p{3.45em}|}
    \hline
        NPU Version & \centering \textbf{A} & \centering \textbf{B} & \centering \textbf{C} & \centering \textbf{D} & \parbox{3.45em}{\centering \textbf{E}} \\\hline
        Deployment Year & \centering 2017 & \centering 2018 & \centering 2020 & \centering 2023 & \parbox{3.45em}{\centering N/A} \\\hline
        Technology & \centering 16nm & \centering 16nm & \centering 7nm & \centering 7nm* & \parbox{3.45em}{\centering 4nm} \\\hline
        Frequency (MHz) & \centering 700 & \centering 940 & \centering 1050 & \centering 1750 & \parbox{3.45em}{\centering 2000} \\\hline
        SA Width & \centering 128 & \centering 128 & \centering 128 & \centering 128 & \parbox{3.45em}{\centering 256} \\\hline
        VU Config. & \multicolumn{5}{c|}{8$\times$128 SIMD Unit} \\\hline
        \# of SAs/VUs & \centering 2/4 & \centering 4/4 & \centering 8/4* & \centering 8/6* & \parbox{3.45em}{\centering 8/8} \\\hline
        SRAM Size (MB) & \centering 32 & \centering 32 & \centering 128 & \centering 128* & \parbox{3.45em}{\centering 256} \\\hline
        HBM Type & \centering HBM2 & \centering HBM2 & \centering HBM2 & \centering HBM2e & \parbox{3.45em}{\centering HBM3e} \\\hline
        HBM BW (GB/s) & \centering 600 & \centering 900 & \centering 1200 & \centering 2765 & \parbox{3.45em}{\centering 7400} \\\hline
        HBM Size (GB) & \centering 16 & \centering 32 & \centering 32 & \centering 95 & \parbox{3.45em}{\centering 192} \\\hline
        ICI BW/link (GB/s) & \centering 62 & \centering 70 & \centering 50 & \centering 100 & \parbox{3.45em}{\centering 150} \\\hline
        ICI Config. & \multicolumn{2}{c|}{4 links/chip, 2D Torus} & \multicolumn{3}{c|}{6 links/chip, 3D Torus} \\\hline
    \end{tabular}
    \label{tab:npu_specs}
\end{table}

\section{Power Gating Opportunities in NPUs}
\label{sec:design_pg:study}

To understand how to enable power gating in NPUs, we study the energy consumption of NPU chips running recent ML workloads (see \Cref{tab:workloads}).
We use a production-level simulator that reports the execution time and static/dynamic energy (i.e., energy consumed due to static/dynamic power) of core components (i.e., SA, VU, SRAM, HBM, and ICI) in each tensor operator.
The simulator is validated with prior studies and real NPU chips (see \mbox{\S\ref{sec:design:impl}}).
Our study covers four NPU generations derived from Google TPUs (see \Cref{tab:npu_specs}).
For a fair comparison across generations, we account for the performance differences as follows.
For each workload, we first profile its performance using the default batch size with the minimum number of NPU-D chips.
We use 1/5 of this performance (i.e., 5 times of latency for inference or 1/5 of throughput for training) as the service-level objective (referred to as 1$\times$ SLO)~\cite{dynamollm:hpca25}.
For each NPU version, we use the most energy-efficient SLO-compliant configuration (see \Cref{tab:workload_config}).
We assume a 60\% duty cycle for NPU chips~\cite{meta:sustainable_ai:mlsys2022} and a power usage efficiency (PUE) of 1.1~\cite{google_pue}.

\begin{figure*}[t]
    \centering
    \includegraphics[width=\linewidth]{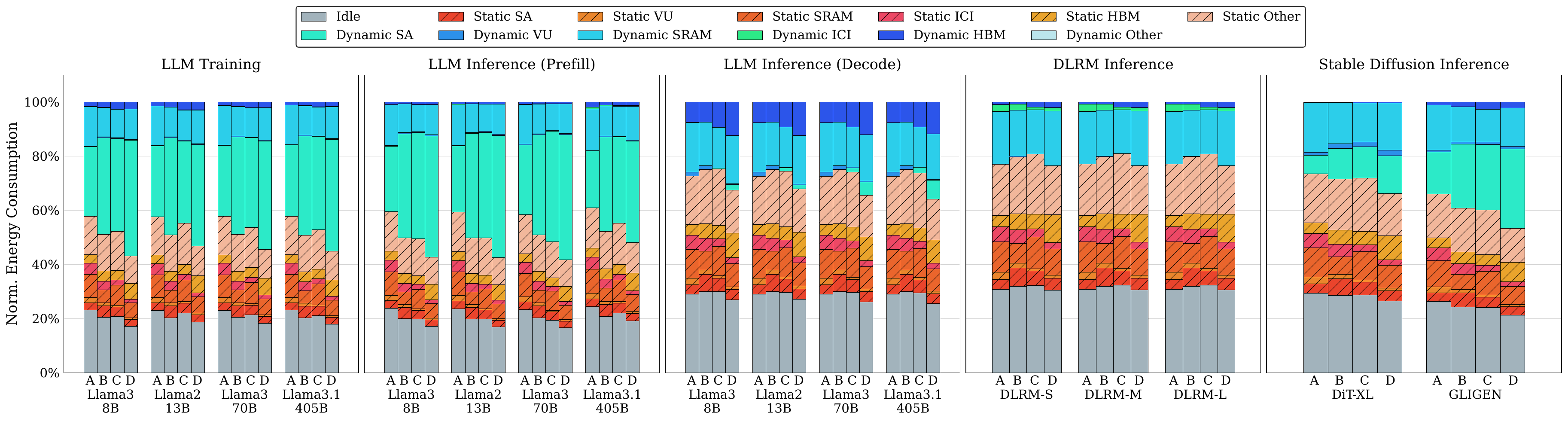}
    \vspace{-4ex}
    \caption{Energy consumption breakdown of ML workloads on different NPU generations.}
    \label{fig:energy_efficiency_breakdown}
\end{figure*}

\noindent
\textbf{End-to-end energy consumption.}
As shown in \Cref{fig:energy_efficiency}, the energy efficiency improves across NPU generations due to two factors: (1) more advanced technology nodes improve the compute and memory/ICI efficiency (FLOPs/W and GB/s/W) , and (2) the larger HBM capacity allows workloads to execute on fewer chips.

However, regardless of the NPU generation, there is a significant energy waste due to static power consumption. We break down the static/dynamic energy of each component of an NPU chip in \Cref{fig:energy_efficiency_breakdown}.
When the chip is powered on but out of its duty cycle (``Idle'' portion in \mbox{\Cref{fig:energy_efficiency_breakdown}}), static power dominates the energy consumption.
This accounts for 17\%--32\% of total energy, which can be substantially reduced as most circuits on the chip are inactive and can be power-gated.
When the chip is busy (excluding the ``Idle'' portion in \mbox{\Cref{fig:energy_efficiency_breakdown}}), static power accounts for 30\%–72\% of energy.
Our findings align with the broader trend and prior studies showing that, as the feature size shrinks, leakage power will continue to be a major contributor to the total chip power despite modern technology nodes such as FinFET or GAA-FET~\mbox{\cite{Mutschler2017,Mutschler2018,Lovati2023,7nmFinFETcharac2014,7nmFinFETvsCMOS2015,accelwattch:micro21}}.




Next, we analyze the per-component utilization to quantify the potential energy savings of power gating and choose the best-fit power-gating strategy.
Unless otherwise specified, the numbers reported below exclude the idle portion in \Cref{fig:energy_efficiency_breakdown}.






\begin{figure}
    \centering
    \includegraphics[width=\linewidth]{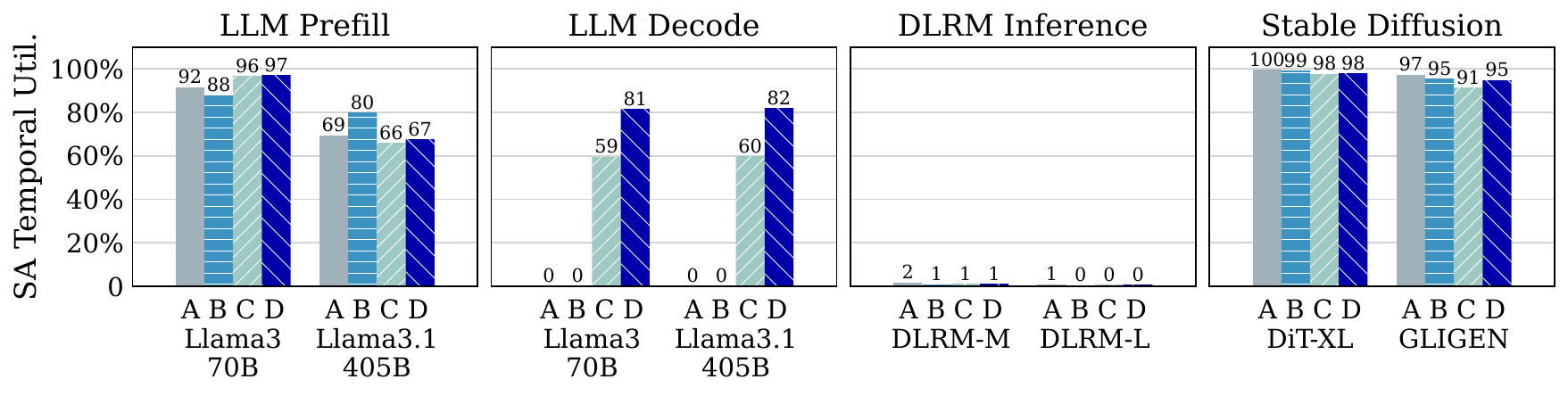}
    \vspace{-4ex}
    \caption{SA temporal utilization (numbers $\leq$0.1\% are rounded to 0).}
    \label{fig:sa_temp_util}
\end{figure}

\begin{figure}
    \centering
    \includegraphics[width=\linewidth]{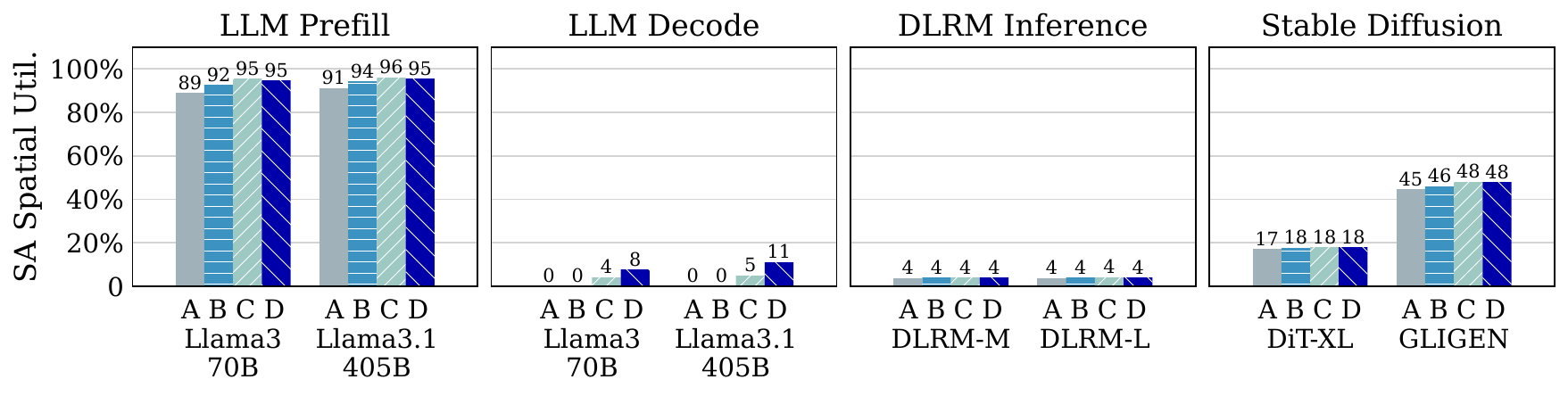}
    \vspace{-4ex}
    \caption{SA spatial utilization quantified by the achieved FLOPs over the theoretical peak FLOPs during SA active time.}
    \label{fig:sa_spatial_util}
\end{figure}

\begin{figure}
    \centering
    \includegraphics[width=\linewidth]{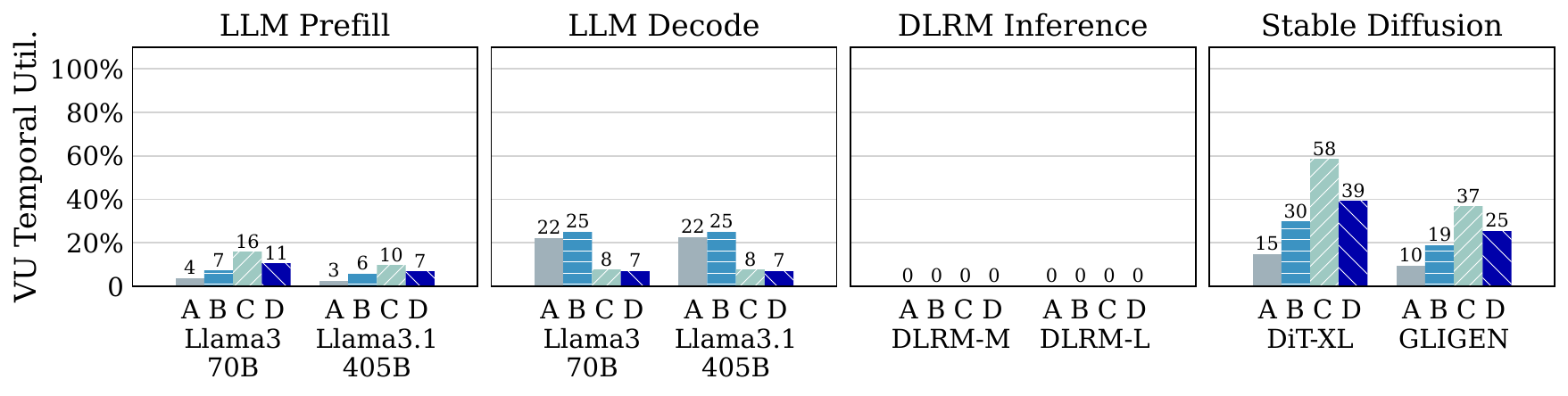}
    \vspace{-4ex}
    \caption{VU temporal utilization.}
    \label{fig:vu_temp_util}
\end{figure}

\begin{figure*}
    \centering
    \includegraphics[width=0.9\linewidth]{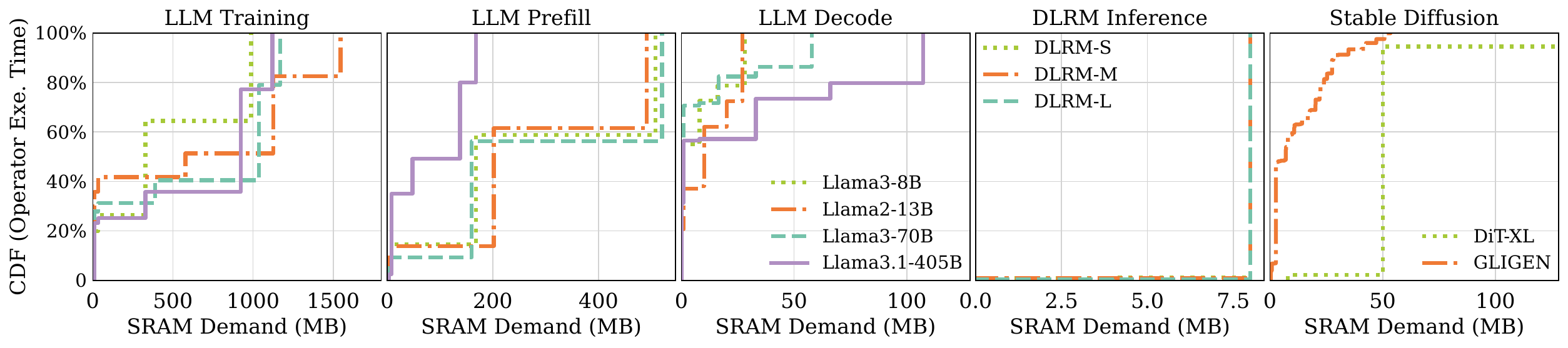}
    \vspace{-1ex}
    \caption{Distribution of SRAM size demands of tensor operators. We only show the results for NPU-D due to space limitations.}
    \label{fig:vmem_size_demand}
\end{figure*}

\noindent
\textbf{Systolic array.}
The SAs contribute 8\%--14\% (10.4\% on average) of static energy.
\Cref{fig:sa_temp_util} shows the SA temporal utilization (quantified by the active cycles).
We do not show LLM training due to space limitations. It has a similar pattern to LLM prefill (both consist of compute-intensive LLM layers with large sequence lengths).
The SA temporal utilization is high for LLM training, LLM prefill, and stable diffusion, as they are compute-bound and involve intensive matrix multiplications (MatMuls) and convolutions (Convs).
The utilization is low for LLM decode and DLRM inferences for two reasons.
First, these workloads are memory- or network-bound.
Second, due to the auto-regressive nature of LLM decode, the embedding tensors are typically too small to amortize the systolic array warm-up latency, so MatMuls may be mapped to the VU.
As the SA underutilization is caused by the intrinsic ML model architectures, it cannot be easily solved by software optimizations like operator fusion and reordering (this observation holds for other components as well, as discussed below).
Ideally, by power gating SAs at component granularity, we can save up to 100\% static SA energy for non-compute-bound workloads and up to 34\% for SA-intensive ones, accounting for 0.8\%--10.1\% of total energy.

\Cref{fig:sa_spatial_util} analyzes the SA spatial utilization.
We focus on LLM prefill and stable diffusion, as they have high temporal SA utilization.
LLM prefill utilizes most processing elements (PEs) in an SA since it processes a large number of input tokens, and the tensors are large enough to saturate the SA.
Stable diffusion models cannot utilize all PEs: DiT-XL has an attention head size of 72, which is smaller than the SA width (128); GLIGEN uses a U-Net architecture that shrinks the image size and attention head size in deeper layers.
This issue will be more severe as future NPU generations use larger SAs~\cite{cloudtpu:google}.
Ideally, with power gating at PE granularity, we can save up to 83\% static SA energy (4.2\% of total energy).







\noindent
\textbf{Vector unit.}
The VUs contribute 1.9\%--5.6\% of static energy.
\Cref{fig:vu_temp_util} shows the VU temporal utilization of ML workloads.
The VU utilization is low (below 60\% for all workloads) due to two reasons.
First, in SA-bound operators (MatMul and Conv), VUs are mainly used to post-process SA outputs.
The VU will be idle waiting for SA outputs despite possible operator fusion opportunities (e.g., MatMul followed by ReLU activation)~\cite{v10:isca23,neu10:micro24,neucloud:hotos23}.
Second, non-SA operators (e.g., Softmax, LayerNorm) are typically memory-bound due to their low arithmetic intensity, so VUs will be waiting for HBM accesses.
The VU spatial underutilization is not a concern due to its small size.
Ideally, power gating VUs can save up to 4.0\% of total energy.

\noindent
\textbf{On-chip SRAM.}
SRAM accounts for 15.4\%--24.4\% (20.9\% on average) of static energy.
It has high temporal utilization, as it serves both compute units (SAs/VUs) and data transfers (HBM/ICI).
The spatial utilization (SRAM capacity usage) depends on the SRAM demands of tensor operators.
We quantify the demand using the minimum tile size that maximizes the on-chip data reuse. For streaming operators (e.g., elementwise) whose data reuse is not affected by tile size, we use the minimum tile size that hides the HBM latency. We fuse as many consecutive operators as possible when they are small enough to fit entirely into the 128MB SRAM.

\Cref{fig:vmem_size_demand} shows the SRAM demands of ML workloads.
Compute-bound workloads like LLM training, LLM prefill, and stable diffusion require large tile sizes to achieve high arithmetic intensity.
Memory/ICI-bound workloads like LLM decode and DLRM intrinsically have low arithmetic intensity, so a larger tile size will not help further improve the on-chip data reuse. For example, the SRAM demand does not exceed 8MB for DLRM, meaning that we can save at least 10.9\% of total energy by power gating the unused capacity.


\begin{figure}
    \centering
    \includegraphics[width=\linewidth]{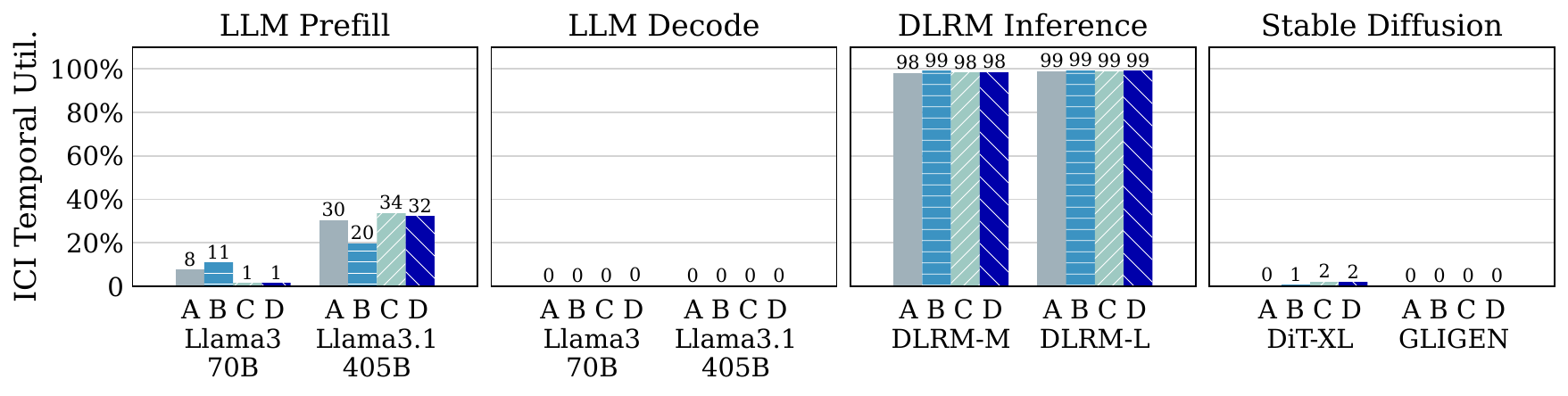}
    \vspace{-4ex}
    \caption{ICI temporal utilization.}
    \label{fig:ici_temporal_util}
\end{figure}

\noindent
\textbf{Inter-chip interconnect.}
The ICI controller \& PHY account for 5.3\%--12.0\% of static energy.
They are active only during collective operators, such as AllReduce, AllToAll, and P2P Send/Recv~\mbox{\cite{megatron-lm,DLRM19}}. Collective operators are typically invoked per layer or microbatch, leaving substantial idle intervals between them. During these intervals, non-collective operators perform local computations (e.g., MatMul and Softmax) without inter-chip communications, allowing the ICI controller \& PHY to be fully power-gated. \mbox{\Cref{fig:ici_temporal_util}} shows that non-collective operators account for 1\%--100\% (67\% on average) of the total execution time.
Ideally, power gating ICI components can save 1.3\% of total energy on average.

\begin{figure}
    \centering
    \includegraphics[width=\linewidth]{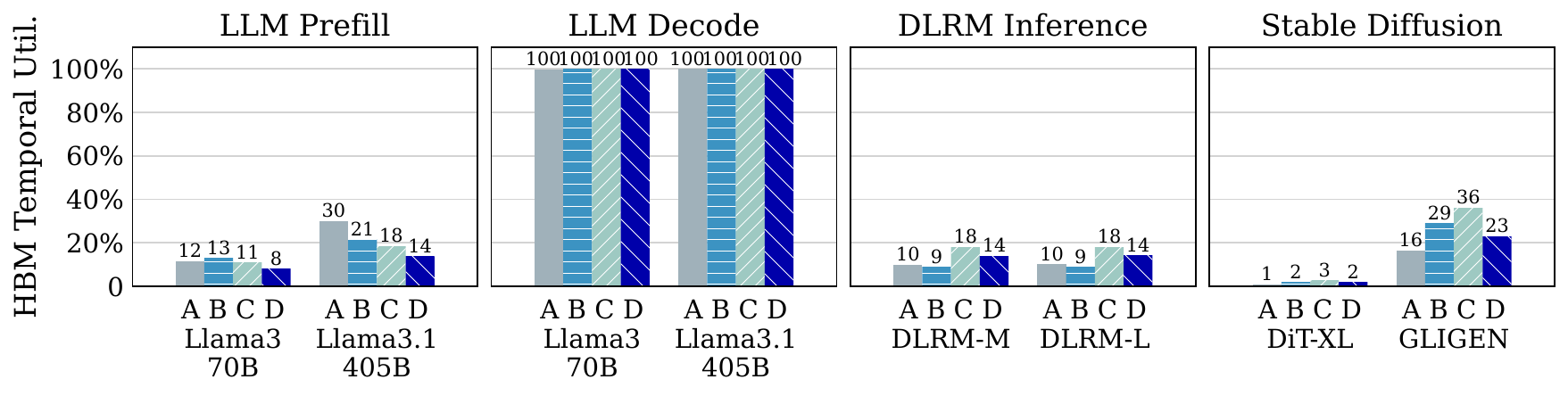}
    \vspace{-4ex}
    \caption{HBM temporal utilization.}
    \label{fig:hbm_temporal_util}
\end{figure}

\noindent
\textbf{HBM controller \& PHY.}
The HBM controller \& PHY account for 9.0\%--22.4\% (12.8\% on average) of static energy.
For compute-intensive workloads (e.g., LLM training/prefill and stable diffusion), HBM is idle for 64\%--99\% (85\% on average) of the total time as shown in \Cref{fig:hbm_temporal_util}. During the idle intervals, the memory controller \& PHY can be power-gated, but they must still periodically (e.g., every 3.9 $\mu$s~\cite{hbm_ip:amd}) perform DRAM refreshes.
This feature is supported by the power-down/self-refresh mode in today's HBM controllers~\cite{hbm_ip:atria,hbm_ip:rambus,hbm_ip:amd}.
Ideally, power gating the HBM controller \& PHY can save up to 12\% (5.3\% on average) of total energy.

\noindent
\textbf{Other components.}
Peripheral components such as chip management, control logic, PCIe controllers, and miscellaneous datapaths account for 39.1\%--45.8\% of static energy.
As they are critical to the regular operation of the chip, we do not aim to power-gate them.

\section{\pname{} Design and Implementation}
\label{sec:design}






We design \pname{} with the following goals:
\begin{itemize}[leftmargin=*]
    \item \textbf{Fine-grained power mode control:} To maximize static power saving, we need to enable the flexibility to partially power-gate a component, such as a portion of an SA or the SRAM.
    \item \textbf{Precise power gating decision:} We need to accurately capture the idle periods to decide the best timing for power gating.
    \item \textbf{Minimal overhead:} We should minimize the performance overhead for ML workloads and the hardware cost of \pname{}.
    \item \textbf{Flexibility:}
    We should provide developers with the flexibility to choose between HW- and SW-managed power gating.
\end{itemize}



\pname{} first enables hardware support for power-gating each component on an NPU chip.
By default, \pname{} supports hardware-managed power-gating policies ($\S$\ref{sec:design_pg:hw}).
To fully exploit the predictability of ML workloads, \pname{} extends the NPU ISA to expose power management functions to the software ($\S$\ref{sec:design_pg:isa}).
The compiler can perform static analysis on the NPU program to precisely capture the idle periods of execution units and the SRAM capacity demand. It inserts power management instructions into the NPU program to implement software-defined power gating policies ($\S$\ref{sec:design_pg:sw}).



\subsection{Hardware Support for NPU Power Gating}
\label{sec:design_pg:hw}

We first present \mbox{\pname{}}'s default hardware-managed power-gating support for each component.
Then, we discuss how the NPU core tracks the power states of all components to enforce dependencies when an instruction needs to wait for a component to wake up.

\begin{figure}
    \centering
    \includegraphics[width=\linewidth]{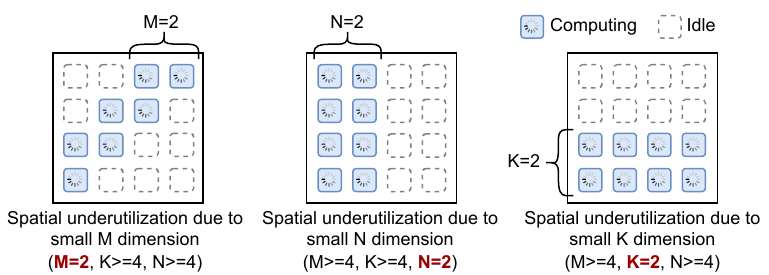}
    \vspace{-4ex}
    \caption{SA spatial underutilization caused by small $M$/$N$/$K$ in a matrix multiplication of shape $[M,K]\times [K,N] \to [M,N]$. We assume a weight-stationary dataflow following the TPU architecture.}
    \label{fig:sa_underutil_mnk}
\end{figure}

\noindent
\textbf{Spatially power-gated systolic array.}
As discussed in $\S$\ref{sec:design_pg:study}, we need to power gate at the PE level to exploit the spatial underutilization.
\Cref{fig:sa_underutil_mnk} illustrates the three cases in which a MatMul operator of shape $[M,K]\times [K,N] \to [M,N]$ can underutilize the SA. To utilize all PEs, all of $M$, $N$, and $K$ must be larger than SA width.
When $N$ or $K$ is smaller than SA width, some rows/columns of PEs will be idle for the entire operator and should be completely powered off.
When $M$ is smaller than SA width, all PEs still need to hold the weights. 
However, as the input data flows through the SA diagonally, a PE is active only when the input data passes through.
The entire PE, except the weight register, can be power-gated otherwise. 

To realize the above power-gating strategy, a na\"ive way is to employ an idle-detection mechanism for each PE.
However, this will incur significant performance overhead for the first case in \Cref{fig:sa_underutil_mnk}, as each PE will suffer at least a cycle of wake-up delay, doubling the execution time when $M$ is smaller than SA width.
Even worse, this cannot support the second and third cases in \Cref{fig:sa_underutil_mnk}, as the compiler will pad zero weight values to fit SA width, and the hardware is unaware whether $N$ or $K$ is smaller than SA width.
Simply adding zero-weight detection in each PE is still insufficient since the PE may still need to pass input data and partial sum to other PEs and cannot be completely power-gated.
Since the PE idleness can change in every cycle, it is also impractical for the software to control the power modes of all PEs in every cycle.
To address these challenges, we design a row/column-wise power-gating mechanism based on non-zero weight detection and leverage the diagonal dataflow to enable cycle-level PE power-gating.

\begin{figure}
    \centering
    \includegraphics[width=\linewidth]{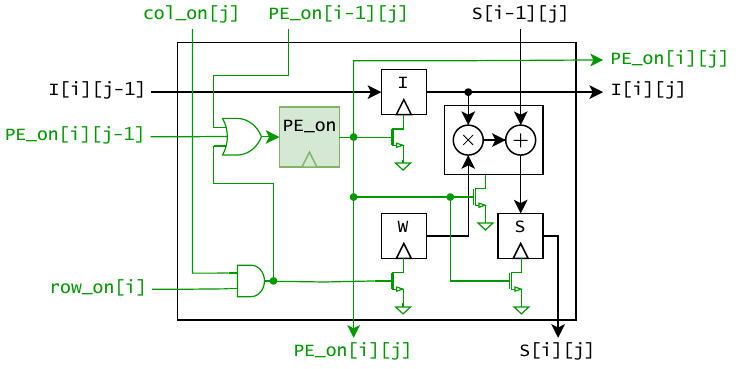}
    \vspace{-4ex}
    \caption{PE architecture for power gating assuming NMOS Gated-Vdd~\cite{gatedvdd:islped00}. \texttt{i}/\texttt{j} is the row/column index of the PE. All input signals are shown on the left/top, and all outputs are shown on the right/bottom.}
    \label{fig:power_gating_PE}
\end{figure}

\noindent
\textul{PE architecture for power gating.}
We define three power modes for a PE: \texttt{ON}, \texttt{W\_on}, and \texttt{OFF}.
\Cref{fig:power_gating_PE} shows the PE architecture.
By default, the PE is \texttt{OFF} and completely power-gated.
The PE receives four power-gating control signals. \texttt{row\_on}/\texttt{col\_on} specifies which rows/columns of PEs should be active. 
When both signals are high, the PE is put into \texttt{W\_on} mode, in which only the weight register (\texttt{W}) is powered on.
The two \texttt{PE\_on} signals will be asserted by the left/top PEs as the input data arrives. 
When one of them is high, the PE will turn into \texttt{ON} mode, waking up all registers and the ALU.

\begin{figure}
    \centering
    \includegraphics[width=\linewidth]{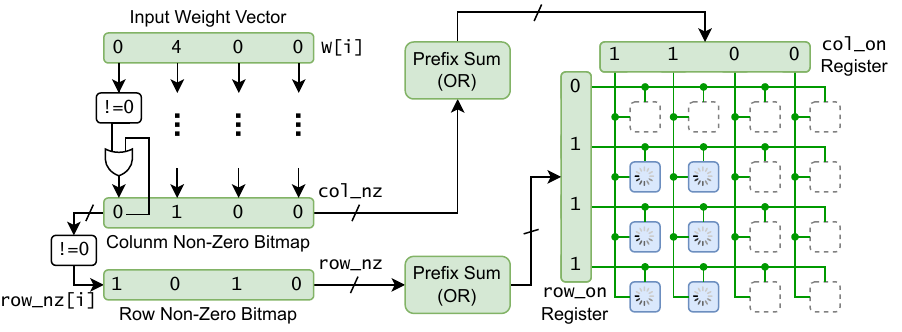}
    \vspace{-4ex}
    \caption{Row/column-wise power-gating control logic for SA based on zero-weight detection. This enables power-gating for the underutilized PEs along the $N$/$K$ dimension.}
    \label{fig:SA_row_col_PG_control}
\end{figure}

\noindent
\textul{Row/column-wise power gating.}
To exploit the $N$/$K$ underutilization, we enable row/column-wise power gating by detecting zero weight values, as shown in \Cref{fig:SA_row_col_PG_control}.
As weight values are being pushed into the SA row by row, we detect non-zero values to generate the column and row bitmaps (\texttt{col\_nz} and \texttt{row\_nz}).
The bitmaps indicate which row or column contains at least one non-zero weight value.
As the input data flows diagonally in the SA, a row/column can be put into OFF mode if (1) it contains only zero weights and (2) all rows/columns to its top/right contain only zero weights.
This condition is detected by performing a prefix sum on the \texttt{col\_nz} and \texttt{row\_nz} bitmaps.
For example, if \texttt{col\_nz=4'b0100}, after the prefix sum, the \texttt{col\_on} bitmap will be \texttt{4'b1100}: column 0 is ON despite its zero weights, as it needs to pass input data to column 1.

\begin{figure}
    \centering
    \includegraphics[width=0.95\linewidth]{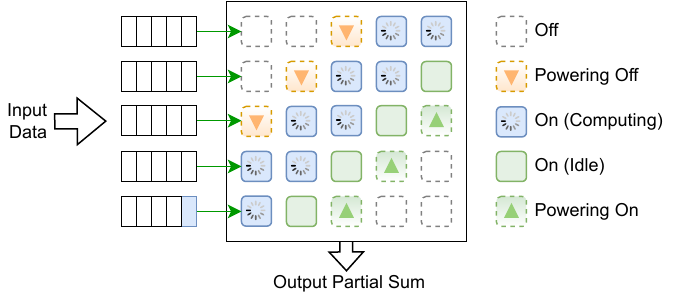}
    \vspace{-2ex}
    \caption{Dataflow of power gating systolic array when the $M$ dimension is underutilized.}
    \label{fig:PG_SA_dataflow}
\end{figure}

\noindent
\textul{Dataflow of a diagonally power-gated SA.}
\Cref{fig:PG_SA_dataflow} shows how the PEs will be power-gated and woken up following the diagonal dataflow, which exploits the $M$ underutilization.
The SA maintains a queue per row to stage the input data.
When the data arrives at the front of the queue, the \texttt{PE\_on} signal to the first PE will be asserted.
Then, the input data can flow into the \texttt{I} register.
As the computation starts on the first PE, the \texttt{PE\_on} signals will be propagated diagonally to wake up other PEs in advance.
Once the queue is empty, the \texttt{PE\_on} signal will be de-asserted for the first PE in the row, and this inactive signal will propagate to put PEs into \texttt{W\_on} mode.
By overlapping the computation and the PE wake-up delay, we can reduce the exposed SA wake-up delay to the single-PE wake-up delay.
The \texttt{PE\_on} signal only adds one extra bit of wire between neighboring PEs compared to the original 16-bit input or 32-bit partial sum. It incurs little area overhead and does not introduce timing issues (see our synthesized results in \mbox{\S\ref{sec:design:impl}}).

\noindent
\textbf{Vector unit.}
The NPU architecture leverages a statically scheduled VLIW ISA to simplify hardware design~\cite{tpudesign:google:ieeemicro21}.
This implies that the exact idle periods of a VU depend on workload characteristics and compiler optimizations (e.g., operator fusion, instruction scheduling).
As there is no dynamic instruction scheduling in the hardware, it will require significant engineering effort and hardware overhead to support an accurate idle period predictor in the hardware.
\pname{} uses a simple idle-detection state machine to power-gate a VU during its idle periods with the best effort.
To minimize the performance impact, we set the idle-detection threshold to at least 8 cycles to avoid blocking the SA execution, as each VU can process $8\times$ output elements from an SA in each cycle.
We leave it to the software to enable precise power-gating for the VUs (see $\S$\ref{sec:design_pg:sw}).

\noindent
\textbf{Segment-wise power-gated SRAM.}
\pname{} divides the SRAM into equally sized segments and enables per-segment power-gating.
By default, the segment size is the same as the vector register size (4KB in our NPU).
Each segment has three power states: \texttt{ON}, \texttt{SLEEP}, and \texttt{OFF}.
The \texttt{OFF} mode leverages Gated-Vdd~\cite{gated-vdd} to minimize leakage power without retaining data.
However, as the SRAM is used as a software-managed scratchpad memory, it is impossible for the hardware to detect which segment is never used and contains no useful data.
Hence, \pname{} also supports a \texttt{SLEEP} mode, which switches the supply voltage of the SRAM cells to a lower value to reduce leakage power while retaining data~\cite{drowsy_cache:isca02}.

Most operators in an ML model follow a tile-by-tile computation pattern, and previous tiles will not be accessed again in the near future.
In the worst case, a VU can consume 8$\times$128$\times$3$\times$fp32=12KB per cycle, so it will take 1,365 cycles for 8 VUs to consume all data in a 128MB SRAM.
Since most operators typically have a higher compute intensity, most segments will be idle for a long period, implying that a simple power-gating policy that periodically puts unused segments into sleep can be employed to exploit the idle periods.
\mbox{\pname{}} employs this simple idle-detection policy for hardware-managed SRAM power-gating.
We leave it to the software to exploit the \texttt{OFF} mode, as the compiler has precise SRAM allocation information to fully power off unused segments (see \mbox{$\S$\ref{sec:design_pg:sw}}).



\noindent
\textbf{HBM/ICI controller \& PHY.}
The HBM is accessed by the DMA engine, and DMA requests in NPUs typically have large sizes.
This is especially true for compute-intensive workloads, which have high HBM idleness and favor a large tile size to improve arithmetic intensity.
Hence, an idle-detection policy is sufficient, as the wake-up delay can be amortized by the long DMA latency.
During the idle period, \pname{} powers off the DMA engine (if it is doing RDMA over ICI) and configures the HBM controller into low-power auto-refresh mode.
Similarly, idle-detection is sufficient for the ICI controller \& PHY, as they are only used in collectives and have long idle periods.

\noindent
\textbf{Power state management in NPU core pipeline.}
In \mbox{\pname{}}, a power-gated component is handled as a structural hazard when an instruction needs to stall until it wakes up.
Taking the Google TPU design as an example, the NPU core pipeline manages structural hazards between different VLIW instruction bundles with standard hold conditions on the instructions~\mbox{\cite{tpudesign:google:ieeemicro21}}: an instruction cannot be dispatched unless all required components are ready. In \mbox{\pname{}}, a power-gated component will be set as ``not ready'' to prevent any operations from being dispatched to it. An operation will trigger a wake-up signal to the corresponding component. The wake-up signal has no effect if the component is already powered on. Otherwise, it will wake up the component and set the ready bit.
As each component has its own ready bit, different components can be powered on/off independently without causing synchronization issues.
By leveraging the existing dependency tracking mechanism, \mbox{\pname{}} minimizes intrusive modifications to the NPU pipeline.


\subsection{ISA Extension for NPU Power Management}
\label{sec:design_pg:isa}

\begin{figure}
    \centering
    \includegraphics[width=0.9\linewidth]{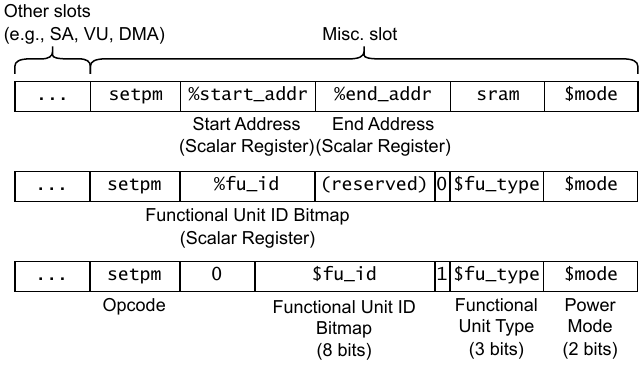}
    \vspace{-2ex}
    \caption{Format of NPU power management instructions. ``\texttt{\%}'' represents a register operand. ``\texttt{\$}'' represents an immediate value. The exact bit width of a field depends on the NPU specification. This figure assumes an NPU with 8 SAs and 8 VUs, so the \texttt{fu\_id} is 8 bits.}
    \label{fig:isa_extension}
\end{figure}

To support software-managed power gating, \pname{} defines three power modes for each component: \texttt{on}, \texttt{auto}, and \texttt{off}.
For SRAM, it also defines a \texttt{sleep} mode.
The \texttt{auto} mode is enabled by default, which employs HW-managed policies to transparently control the power gating of all components.
These power modes offer the flexibility to choose between HW- and SW-managed power gating.
As the HW-based approaches work immediately without software support, this helps maintain compatibility with legacy programs, and serves as a transitional solution while new SW-based power optimizations are being developed for evolving workloads.

We introduce a new \texttt{setpm} (set power mode) instruction to configure a component into the \texttt{auto}/\texttt{on}/\texttt{off}/\texttt{sleep} mode.
\Cref{fig:isa_extension} shows its semantics.
\texttt{setpm} is encoded in the miscellaneous (misc.) slot in a VLIW instruction bundle.
Since components have different power-gating granularity, we define three variants of \texttt{setpm}.

The first variant is used for SRAM and is indicated by setting the \texttt{fu\_type} field to \texttt{sram}.
It takes two scalar register operands that specify the start and end addresses of a contiguous region in the SRAM.
This enables the software to ``shrink'' the SRAM capacity with a single instruction based on the ML workload demand.

The second and third variants are used for other components.
As there are multiple SAs/VUs, we use a bitmap (\texttt{fu\_id}) to indicate which SA/VU should be affected by a \texttt{setpm}.
The bitmap can be either from a scalar register or an immediate field.
For example, \texttt{setpm 0b1011,vu,off} will power-gate VU 0, 1, and 3.
Using a bitmap requires only a few extra bits than specifying the functional unit index, which is negligible compared to the entire VLIW instruction (322 bits on TPUv2~\mbox{\cite{tpudesign:google:ieeemicro21}}).
The bitmap allows a single \texttt{setpm} to control multiple functional units.
This reduces the number of \texttt{setpm} instructions, which is desirable for two reasons.
First, \texttt{setpm} instructions need to be fetched and decoded, which consumes dynamic energy or affects performance.
Second, since we can only issue one \texttt{setpm} per cycle (with one misc. slot), configuring multiple functional units in parallel reduces the wake-up delay. 

\subsection{Software-Managed NPU Power Gating}
\label{sec:design_pg:sw}

\begin{figure}
    \centering
    \includegraphics[width=\linewidth]{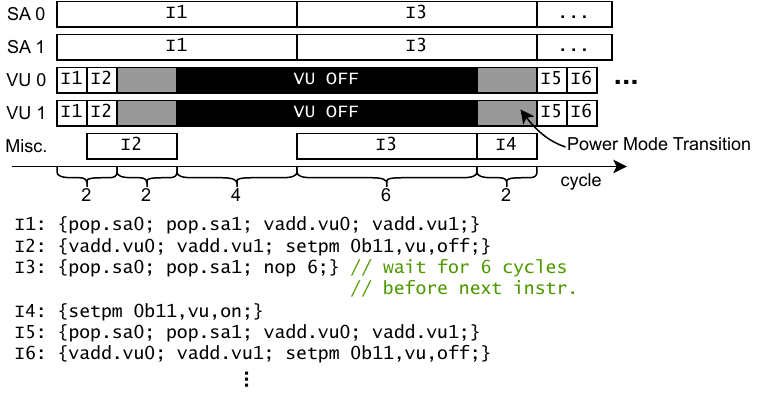}
    \vspace{-5ex}
    \caption{Example code snippet using \texttt{setpm} to power gate VUs (bottom) and corresponding execution timeline (top). The NPU has 2 SAs and 2 VUs. \texttt{push} takes 8 cycles to feed 8$\times$128 elements into an SA. \texttt{vadd} takes 1 cycle on a VU. VU power-on/off delay is 2 cycles.}
    \label{fig:pg_prog}
\end{figure}

To implement software-managed power gating, \pname{} extends the ML compiler to analyze the idle periods of each component and instrument an NPU program with \texttt{setpm} instructions.
As discussed, SAs, HBM, and ICI should use \texttt{auto} mode.
Hence, we focus on the software-managed power gating support for VUs and SRAM.

\noindent
\textbf{ML compiler framework.}
Modern ML compilers, such as XLA~\mbox{\cite{xla}}, PyTorch (via torch.compile)~\mbox{\cite{pytorch2}}, and TensorRT~\mbox{\cite{tensorrt}}, typically assume a static computation graph with known tensor shapes and bounded loops.
This practice enables aggressive compiler optimizations that significantly improve performance and is widely adopted in production.
Workloads with dynamic behaviors still consist of static subgraphs.
For example, in a conditional branch, each branch (typically a subset of layers or modules in the model) may be treated as a static subgraph; variable input tensor shapes (e.g., batch sizes or sequence lengths) can be bucketized, and a static graph is compiled for each bucket.
Static graphs enable accurate component idleness analysis (i.e., no prediction errors in theory) at compilation time, making software-managed power gating practical and effective.

The static graph is lowered by the compiler frontend into an optimized intermediate representation (IR) and passed to the device-specific compiler backend to generate the assembly code.
The backend can be connected to different frontends by taking their IRs (e.g., XLA HLO~\mbox{\cite{xla_backend}} or TorchInductor IR~\mbox{\cite{pytorch2}}) as inputs.
\mbox{\pname{}} integrates the component idleness analysis pass and the \texttt{setpm} instrumentation pass into the backend after the instruction scheduling pass and the SRAM memory allocation pass.

\noindent
\textbf{Component idleness analysis.}
To determine when to power-gate a VU and which portion of SRAM can be power-gated, the ML compiler needs to extract the idle intervals of VUs and the lifetime of each buffer allocated in the SRAM.
It calculates the distances in cycles between consecutive instructions in the same VU slot.
If there exists a DMA operation between two VU instructions, their distance is set to infinity, as the DMA will be at least the HBM latency (a few hundred nanoseconds), which is much longer than the BET of VU.
For SRAM, \pname{} uses the output of the SRAM allocation pass, which includes the lifetime (start/end instruction index), start address, and size of each allocated buffer.
Based on this, \pname{} derives the idle intervals of each 4KB segment.

\noindent
\textbf{BET-based power-gating policy.}
\pname{} uses the break-even time (BET)~\cite{warped_gates:micro13} and the power-on/off delay of each component to decide whether it should be power-gated in an idle interval.
If the idle interval is longer than BET and $2\times$ the power-on/off delay, \pname{} will insert \texttt{setpm} instructions at the start and end of the interval.
Otherwise, an idle interval longer than BET means the dynamic energy spent on powering on/off the component will outweigh the energy saved by power gating. An interval longer than $2\times$ the power-on/off delay means a potential performance overhead.
In these two cases, we should not power gate the component.
The exact value of BET and power-on/off delay depends on the hardware circuit implementation and is provided by the chip manufacturer (see \mbox{\S\ref{sec:design:impl}} for more details in our hardware synthesis results).
The ML compiler takes the BET values as configurable compiler flags.

\Cref{fig:pg_prog} shows an example code snippet from a MatMul operator.
The VUs post-process SA outputs and are only active for 2 cycles out of each 16-cycle period.
\pname{} maximizes the power-gated cycles of VUs (10 cycles in the example) by inserting \texttt{setpm} at the most appropriate timings.
In contrast, the HW-based policy will waste a few cycles to observe idleness before gating the VUs.




\subsection{\pname{} Implementation}
\label{sec:design:impl}

\begin{table}[t]
    \centering
    \caption{Power on/off delays and BETs of each component in our synthesized prototype. ``SA (PE/full)'' refers to a single PE vs. the entire SA. ``SRAM (sleep/off)'' refers to putting a 4KB SRAM segment into sleep mode (retaining data) vs. off mode (losing data).}
    \vspace{-2ex}
    \footnotesize
    \begin{tabular}{|c|c|c|c|c|c|c|c|}
    \hline
         & SA & SA & \multirow{2}{*}{VU} & \multirow{2}{*}{HBM} & \multirow{2}{*}{ICI} & SRAM & SRAM \\
         & (PE) & (full) & & & & (sleep) & (off) \\\hline
        Power On/Off & \multirow{2}{*}{1} & \multirow{2}{*}{10} & \multirow{2}{*}{2} & \multirow{2}{*}{60} & \multirow{2}{*}{60} & \multirow{2}{*}{4} & \multirow{2}{*}{10} \\
        Delay (cycle) &  &  &  &  &  &  &  \\\hline
        BET (cycle) & 47 & 469 & 32 & 412 & 459 & 41 & 82 \\\hline
    \end{tabular}
    \label{tab:BET_synthesis}
\end{table}

\noindent
\textbf{Synthesis results and hardware overhead.}
We implemented the power-gating control logic of each component in RTL and verified the functionality of the RTL modules on a Xilinx Zynq UltraScale+ ZU19EG FPGA SoC.
We synthesized the RTL prototype in Cadence Genus and placed \& routed the design in Cadence Encounter with a 7nm FinFET PDK~\mbox{\cite{asap7pdk}}.
To balance area, performance, and energy, we empirically tune the wake-up delay and the number of gating transistors (i.e., the effective gate length~\mbox{\cite{gatedvdd:islped00}}) for each component, while ensuring that the active-state performance is not affected. \mbox{\Cref{tab:BET_synthesis}} summarizes the wake-up delays and BETs.

For a 128$\times$128 SA, the area overhead of the row/column-wise power gating control logic is under 0.001\%.
For each PE, we tune the number of the gating transistors such that the wake-up delay is within 1 cycle. This incurs 6.36\% area overhead for each PE. This is acceptable as the SAs only account for 10.7\% of the total chip area~\cite{tpuv4i:isca21}, so the spatial SA power-gating technique in \pname{} only adds 0.68\% area to a TPUv4i chip\footnote{We use TPUv4i as a reference, since there is no public floorplan for newer TPUs.}.
Upon a \texttt{setpm} command, the entire SA can be powered on/off in 10 cycles. This only requires a few extra wires with negligible area overhead.
For VUs, we target a 2-cycle wake-up delay, which incurs 0.13\% area overhead on a TPUv4i chip.
For each 4KB SRAM segment, we target a 4/10-cycle wake-up delay for \mbox{\texttt{sleep}/\texttt{off}} mode.
Implementing this for a 128MB SRAM takes an extra 2.5\% chip area.
The HBM controller \& PHY IP already supports low-power mode~\cite{hbm_ip:amd}.
The extra idle-detection logic has negligible overhead.
As we power-gate all ICI-related IPs as a whole, the power-gating logic takes negligible area.
In total, \pname{} incurs 3.3\% hardware area overhead on a TPUv4i chip.

\noindent
\textbf{Simulator methodology.}
We implement \pname{} with a production-level NPU simulator based on the TPU architecture.
It takes the ML model graph and simulates the execution of each component (e.g., SA, VU, SRAM, HBM, ICI) in each tensor operator.

The simulator frontend applies common ML compiler optimizations used in production~\mbox{\cite{xla,tvm}}, such as tiling, operator fusion, and operator reordering.
For each operator, it generates tile-level information, including computation (e.g., pushing/popping a tile from/to an SA, performing computation of a tile on a VU), SRAM access (e.g., reading/writing a tile), and ICI/DMA operations.
The simulator backend models the execution of operators at tile granularity and reports statistics on each component, including the execution time in cycles, memory/ICI traffic, and FLOPs utilization.

We model the per-component power based on McPAT~\mbox{\cite{mcpat:micro09,mcpat-calib:iccad21}} and NeuroMeter~\mbox{\cite{neurometer:hpca21}}.
We first model the area of each component based on microarchitectural parameters (e.g., SA dimensions and SRAM capacity) and feature size.
Then, we model the static/dynamic power based on chip area, frequency, and technology node-specific parameters, including voltage levels and channel lengths.
We combine this power model and the results from the performance simulator to quantify the energy consumption for each component.
To implement \pname{}, we modify the backend to support idle-detection-based power-gating and account for the wake-up delays.
We also modify the frontend to implement the component idleness analysis pass and add power management events to simulate \texttt{setpm} instructions. The idleness analysis and \texttt{setpm} instrumentation passes have a linear time complexity with respect to the number of instructions. They do not incur a noticeable increase in compilation time compared to other compiler passes~\mbox{\cite{hlo_passes}}. The code size inflation due to \texttt{setpm} is negligible since it does not occur frequently compared to other operations (see \mbox{\Cref{fig:eval_num_setpm}}).

\begin{figure}[t]
    \centering
    \includegraphics[width=\linewidth]{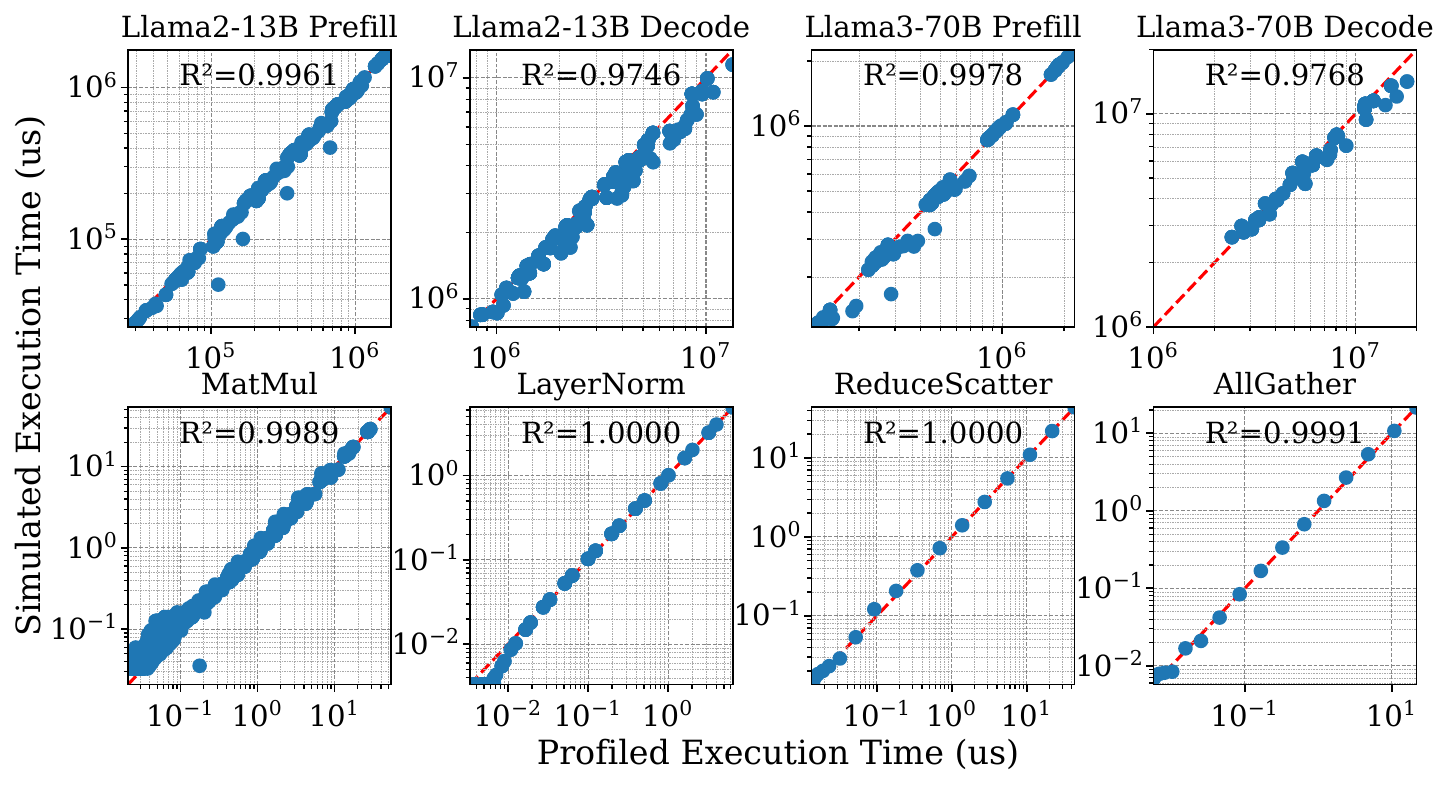}
    \vspace{-4ex}
    \caption{Simulator validation against TPUv4 chips. Due to space limitations, we only show the results of running a single prefill/decode request with 13B/70B Llama models on up to 32 chips (top row). We also show the results of representative operators (bottom row).}
    \label{fig:npusim_validation}
\end{figure}

\begin{figure*}
    \centering
    \includegraphics[width=\linewidth]{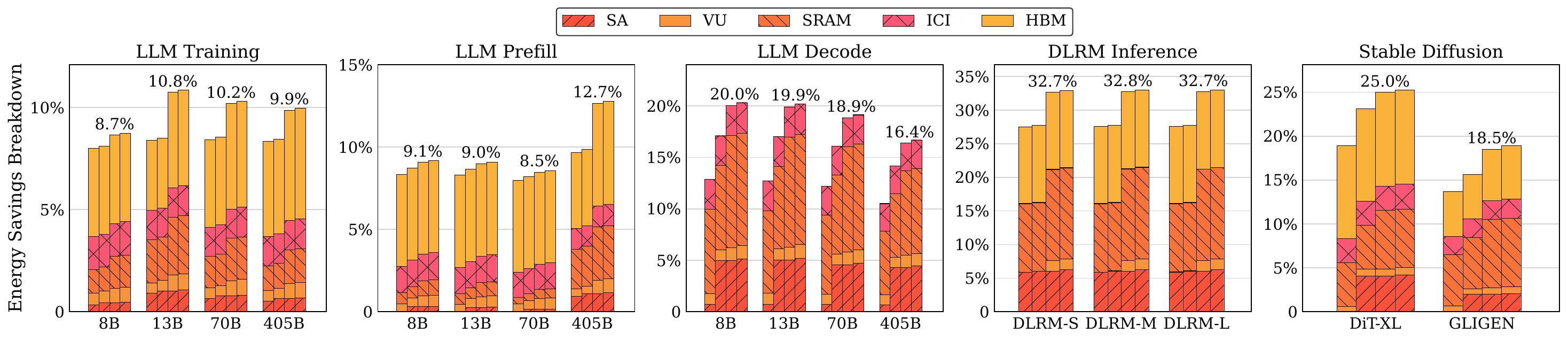}
    \vspace{-4ex}
    \caption{Energy savings normalized to \nopg{}. The bars from left to right are \base{}, \hw{}, \full{}, and \ideal{}.}
    \label{fig:eval_energy}
\end{figure*}

We verified our performance simulator with real TPU chips as well as published TPU literature~\mbox{\cite{neurometer:hpca21,tpuv4:isca23,tpuv4i:isca21,carbon_life_cycle}}.
We run the same ML model graph on both real TPUs and our simulator and validate the end-to-end execution time and the per-component active time for each operator.
\mbox{\Cref{fig:npusim_validation}} shows the simulation accuracy for ML models and operators running on 1 to 32 TPUv4 chips with all possible batch sizes and data/tensor/pipeline parallelism combinations. The Pearson correlation coefficient ($R^2$) of the profiled and simulated results is greater than 0.97 ($R^2=1$ indicates a perfectly linear correlation).
We also validated our chip area and power model against published data~\mbox{\cite{tpudesign:google:ieeemicro21,tpuv4:isca23}}. Our estimated idle/TDP power is within 9\%/5\% for TPUv2 and 10\%/4\% for TPUv3 of published data.
We have open-sourced our simulator (see \Cref{app:artifact}).

\section{Discussion}
\label{sec:discussion}


\noindent
\textbf{Generalizability of \mbox{\pname{}}.}
Most AI chips today employ matrix units (e.g., systolic arrays), vector units, on-chip scratchpad SRAM, off-chip DRAM, and inter-chip links~\cite{h100whitepaper,amazon:trainium,tpuv4:isca23,meta:mtia2i,ipu2,tenstorrent,t10:sosp24,elk:micro25}.
They may also employ specialized functional units for structured sparsity (e.g., sparse tensor cores in GPUs~\mbox{\cite{mishra2021acceleratingsparsedeepneural}}) or specific operators (e.g., embedding cores in TPUs~\mbox{\cite{tpuv4:isca23}}).
For dataflow architectures built of massive interconnected cores~\mbox{\cite{meta:mtia2i,ipu2,tenstorrent}}, each core still consists of those components.
As the underutilization of each component is due to intrinsic characteristics of ML models (see \mbox{\S\ref{sec:design_pg:study}}), there will be abundant opportunities for power gating in these AI chips.
\mbox{\pname{}} offers a systematic solution to power gate those components.
The \texttt{setpm} instruction can be extended to support specialized functional units, and it can also be implemented with non-VLIW ISAs.

\noindent
\textbf{Security implications.}
Power gating amplifies the contrast in the power consumption between idle and active states.
This may potentially increase the risk of power-based side-channel attacks as the power usage pattern is highly correlated with component utilization~\mbox{\cite{intel_amx_power_side_channel}}.
\mbox{\pname{}} allows the software to override the default hardware-managed power gating policies and disable power gating when processing sensitive data, which helps mitigate such attacks.
To completely obfuscate power usage patterns, we can further incorporate software-defined dynamic power management techniques like DVFS and clock gating.
We wish to explore this as future work.

\begin{table}
\caption{The most energy-efficient SLO-compliant configurations for NPU-D chips used on our evaluation.}
\vspace{-2ex}
\footnotesize
\begin{tabular}{|ll|l|l|}
\hline
\multicolumn{2}{|c|}{\textbf{Workload}}                                      & \textbf{\# of Chips} & \textbf{Batch Size} \\ \hline
\multicolumn{1}{|c|}{\multirow{4}{*}{LLM Training/Prefill/Decode}}  & 8B     & 4/1/1                & 32/4/8              \\ \cline{2-4} 
\multicolumn{1}{|c|}{}                                              & 13B    & 4/1/1               & 32/4/4              \\ \cline{2-4} 
\multicolumn{1}{|c|}{}                                              & 70B    & 8/4096/128            & 32/8192/4096          \\ \cline{2-4} 
\multicolumn{1}{|c|}{}                                              & 405B   & 16/256/64           & 32/64/2048            \\ \hline
\multicolumn{1}{|c|}{{DLRM Inference}}                              & S/M/L  & 8/8/8                & 4096/4096/4096      \\ \hline 
\multicolumn{1}{|c|}{\multirow{2}{*}{Stable Diffusion}}             & DiT-XL & 64                  & 8192                \\ \cline{2-4} 
\multicolumn{1}{|c|}{}                                              & GLIGEN & 64                   & 256                  \\ \hline
\end{tabular}
\label{tab:workload_config}
\end{table}

\section{Evaluation}
\label{sec:eval}

Our evaluation shows that:
(1) \pname{} reduces the energy consumption of NPU chips by 8.5\%--32.8\% (15.5\% on average) (\S\ref{sec:eval:energy_saving}) and the average power consumption by 9\%--33\% (15.7\% on average) (\S\ref{sec:eval:power_saving}), with under 0.5\% performance overhead (\S\ref{sec:eval:perf_impact});
(2) \pname{} consistently outperforms the baselines with different threshold voltages, wake-up delays, and NPU chip generations (\S\ref{sec:eval:sens});
(3) When deployed at scale, \pname{} reduces the operational carbon emission of the NPU fleet, which helps improve the optimal NPU device lifespan and facilitates NPU reuse (\S\ref{sec:eval:carbon}).


\subsection{Experimental Setup}
\label{sec:eval:setup}

We evaluate diverse ML workloads in \mbox{\Cref{tab:workloads}} that represent mainstream ML models in production. They cover the most prevalent ML model architectures today, including multi-layer perceptrons (MLPs, common in all models), attentions (Llama, DiT-XL, GLIGEN), large embedding tables (DLRM), and convolutions (GLIGEN). These workloads exhibit diverse characteristics, ranging from compute-intensive, to memory bandwidth-intensive, and ICI-intensive (see the analysis in {\S\ref{sec:design_pg:study}}). We select the most energy-efficient SLO-compliant configuration without power gating (see $\S$\ref{sec:design_pg:study}) for each workload.
We list the configurations in \Cref{tab:workload_config}.
We focus on NPU-D and evaluate other NPU generations in $\S$\ref{sec:eval:sens}.
By default, we set the leakage power of power-gated logic, sleep-mode SRAM, and powered-off SRAM to be 3\%, 25\%, and 0.2\% of the static power in active (ON) mode, respectively~\cite{gatedvdd:islped00,drowsy_cache:isca02}.
The wake-up delay and BET of each component are shown in \Cref{tab:BET_synthesis}.
We vary the leakage power and the wake-up delay in our sensitivity analysis.
Unless otherwise specified, the reported numbers exclude the idle portion in \Cref{fig:energy_efficiency_breakdown}.
We compare the following designs:
\begin{itemize}[leftmargin=*]
    \item \textbf{\nopg{}}: NPU chip without power gating.
    \item \textbf{\base{}}: This variant of \mbox{\pname{}} is configured to resemble conventional hardware-managed techniques to power gate the functional units in CPU/GPU cores. It uses \texttt{auto} mode for all components and disables the spatial power-gating for SAs (i.e., the SA is power-gated at component granularity with idle-detection). The detection window is 1/3 of the BET~\cite{warped_gates:micro13}.
    \item \textbf{\hw{}}: This setup enables our PE-level spatial SA power-gating mechanism on top of \base{}. All components use \texttt{auto} mode with hardware-managed power gating.
    \item \textbf{\full{}}: The full \pname{} design, which adds software-\\managed power-gating on top of \hw{}.
    \item \textbf{\ideal{}}: A roofline with no leakage power in OFF mode, no power-off/on delay, and all idle periods being perfectly power-gated. 
\end{itemize}


\begin{figure*}
    \centering
    \includegraphics[width=\linewidth]{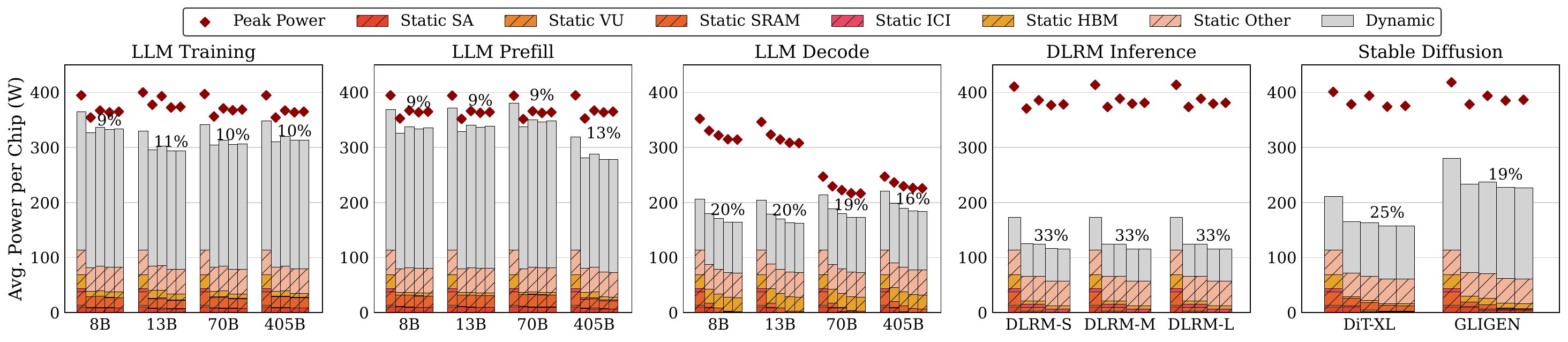}
    \vspace{-4ex}
    \caption{Average/peak power consumption. Peak power is measured as the average power during the execution of the most power-hungry operator. The bars from left to right are \nopg{}, \base{}, \hw{}, \full{}, and \ideal{}.}
    \label{fig:eval_power}
\end{figure*}

\begin{figure*}
    \centering
    \includegraphics[width=\linewidth]{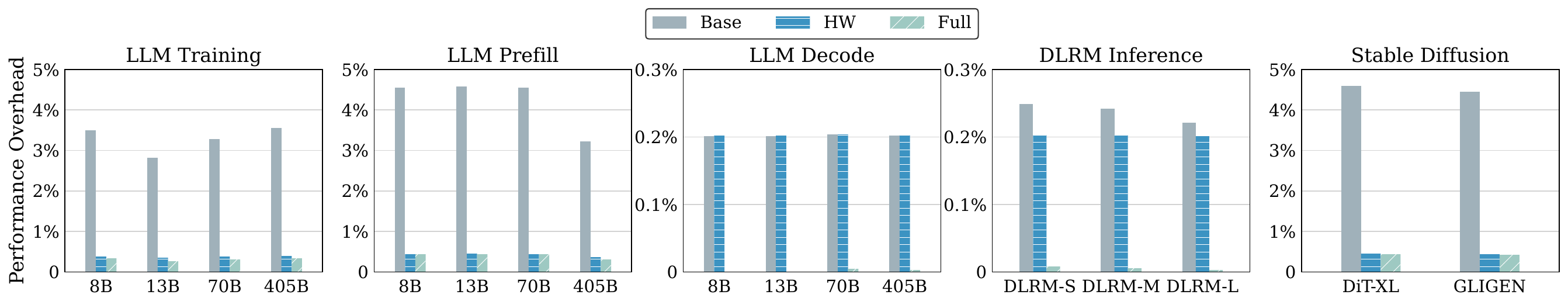}
    \vspace{-4ex}
    \caption{Performance overhead of power gating (normalized to \nopg{}).}
    \label{fig:eval_perf_impact}
\end{figure*}

\begin{figure}
    \centering
    \includegraphics[width=\linewidth]{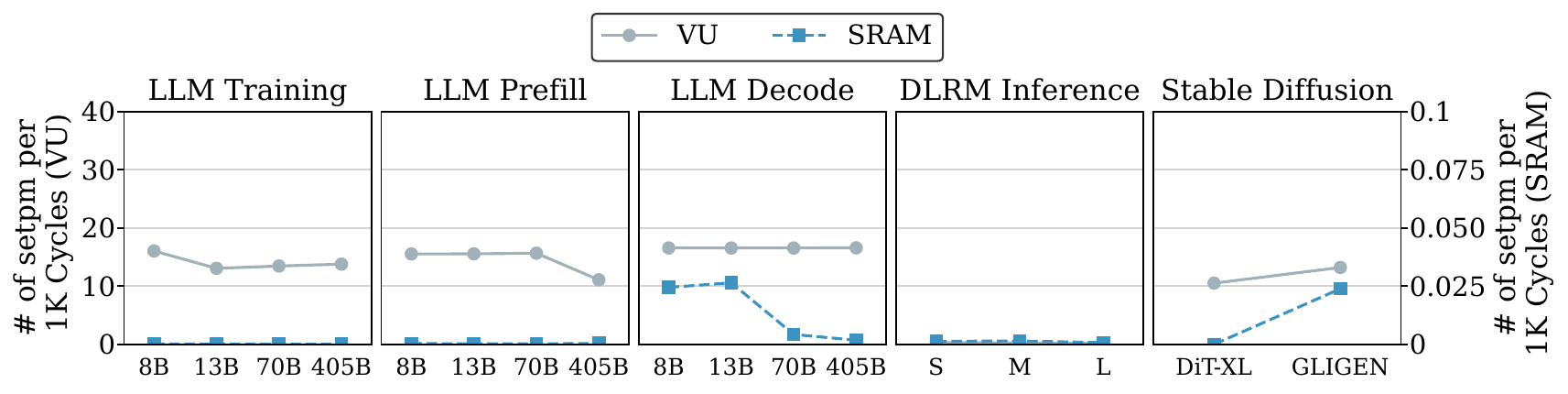}
    \vspace{-4ex}
    \caption{Number of executed \texttt{setpm} instructions per 1,000 cycles in each workload in \mbox{\full{}}.}
    \label{fig:eval_num_setpm}
\end{figure}

\subsection{Energy Savings}
\label{sec:eval:energy_saving}

\Cref{fig:eval_energy} breaks down the energy-saving benefits of different designs compared with \nopg{}.
On average, \full{} reduces the NPU energy consumption by 15.5\% over \nopg{}.

\base{} enables power gating at component granularity.
For example, it power-gates an entire SA when this SA has been temporally underutilized for a certain amount of time.
Thus, for workloads with low SA temporal utilization (e.g., DLRM Inference), \base{} can save up to 27.5\% of energy.

While \base{} only exploits the temporal underutilization of NPU components, \hw{} saves more by further exploiting the spatial underutilization of SAs.
\hw{} power-gates SAs at PE granularity, so underutilized PEs in an active SA can be power-gated individually.
Thus, \hw{} saves up to 22.2\% energy in workloads with low SA spatial utilization (e.g., LLM decode and Stable Diffusion).
The energy savings are less impressive on workloads with high spatial SA utilization (e.g., LLM Training) or low temporal SA utilization (e.g., DLRM Inference).

\full{} achieves 8.5\%--32.8\% energy savings (15.5\% on average) 
for two major reasons.
First, instead of relying on imprecise idle period detection, \full{} immediately shuts down idle components using exact utilization information for each component.
This benefit is most obvious for VUs, whose idle periods are difficult to capture with hardware.
Compared to \base{} and \hw{}, \full{} delivers 3.3$\times$ higher energy savings from VU on average. 
Second, while \full{} powers off unused SRAM segments with compiler knowledge, \base{} and \hw{} must conservatively place unused SRAM cells into sleep mode, which still consumes substantial leakage power. This enables \full{} to deliver 35.5\% more SRAM energy savings than \base{} and \hw{}.
\full{} achieves near-ideal energy saving (within 0.40\% of \ideal{}).
This negligible gap is primarily due to spatial underutilization of VUs and the energy cost of power state transitions.

\subsection{Power Savings}
\label{sec:eval:power_saving}




\Cref{fig:eval_power} shows the average/peak power. The average/peak power of \full{} is 15.7\%/8.3\% lower than that of \nopg{} on average, which matches with the energy savings in \Cref{fig:eval_energy}. Although \base{} occasionally achieves lower power consumption than \hw{} and \full{}, it is still less energy-efficient than \full{} due to its higher performance overhead (see \S\ref{sec:eval:perf_impact}). Conversely, despite slightly higher peak power in certain cases, \full{}'s precise software-informed power-gating decisions ensure it is the most energy-efficient.
Assuming an average cooling cost of \$7/chip-watt~\cite{cooling_cost}, \pname{} can save \$217 per NPU chip on cooling equipment as it reduces the peak per-chip power by 31W on average. This saves \$1.9 million for a TPU pod with 8960 chips~\cite{tpucloud}.



\subsection{Performance Impact}
\label{sec:eval:perf_impact}



\Cref{fig:eval_perf_impact} shows how different power gating designs impact execution performance.
\base{} introduces up to 4.6\% overhead, primarily caused by its power gating policy's inability to hide VU and SA wake-up delays.
SA delays are long for \base{} as it needs to power on the entire SA before execution.
\hw{} allows execution to start after powering on the first PE and overlaps execution with the wake-up of the remaining PEs.
This drastically reduces exposed SA wake-up delay, slashing performance overhead to under 0.6\% on average.
Note that this performance advantage diminishes on memory- or ICI-bound workloads with low temporal SA utilization like LLM Decode and DLRM Inference, as SA wake-up delay is amortized by long SA idle periods.
\full{} leverages the software to wake up the SRAM/VU in advance.
It incurs negligible performance overhead (up to 0.44\%).

\mbox{\Cref{fig:eval_num_setpm}} quantifies the number of executed \texttt{setpm} instructions per 1K cycles for \mbox{\full{}}.
For VUs, the compiler will not insert \texttt{setpm} when the idle interval is shorter than the 32-cycle BET (see \mbox{\S\ref{sec:design_pg:sw}}), implying that the maximum number of \texttt{setpm}s per 1K cycles must be less than $1000/32 \approx 31$. The measured numbers are much lower than this (less than 20 on average), as many VU idle intervals are larger than the BET.
For SRAM, the number of \texttt{setpm}s is negligible.
This is because most SRAM idle intervals come from the unused capacity.
The \texttt{setpm} only needs to be called infrequently when the tile size demand changes (e.g., between operator boundaries or different parts in a fused operator).
The \texttt{setpm}s incur negligible performance overhead as the wake-up delays can overlap with other instructions.
The performance and energy overheads of \texttt{setpm}s are greatly outweighed by the net energy savings.


%
%




\begin{figure}
    \centering
    \includegraphics[width=\linewidth]{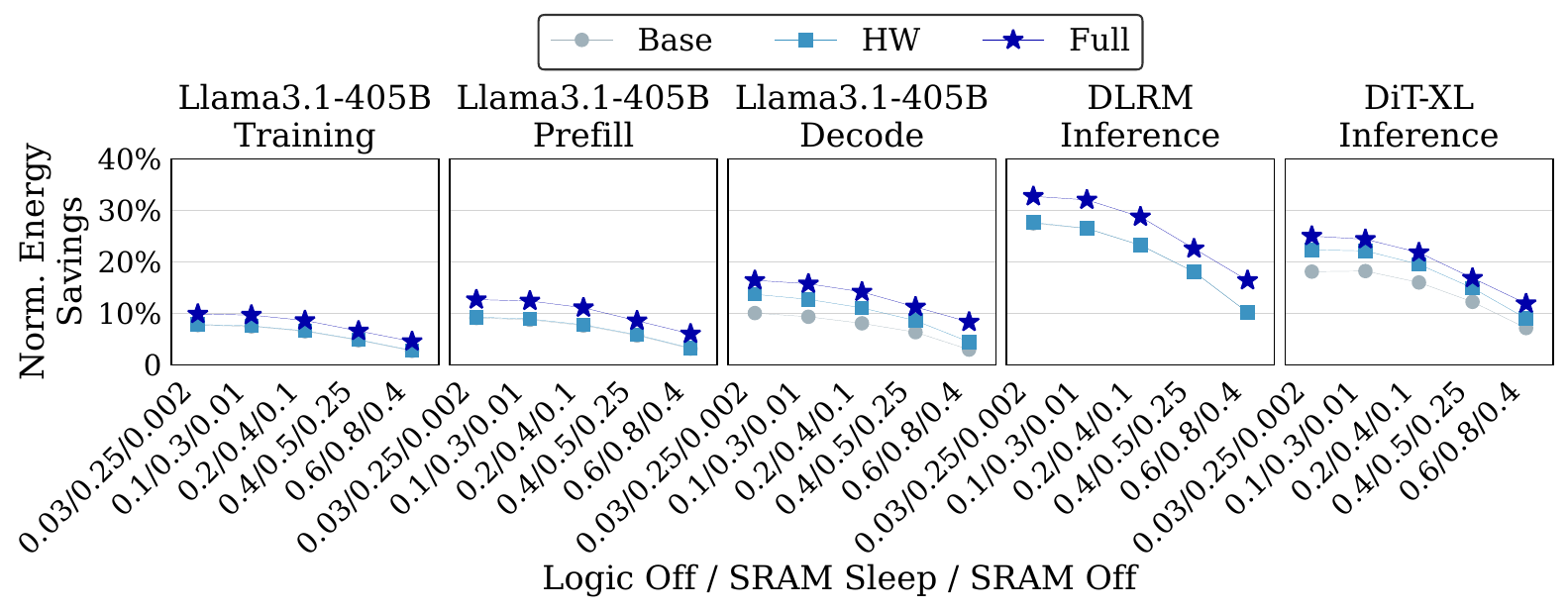}
    \vspace{-4ex}
    \caption{Energy savings of varying power gating threshold voltage. The X-axis labels are the ratios of leakage power of power-gated logic, drowsy SRAM cells, and power-gated SRAM cells, with respect to the static power w/o power gating. For example, 0.03 means the leakage power in the OFF state is 3\% of the ON state static power.}
    \label{fig:eval_vary_Vth}
\end{figure}

\begin{figure}
    \centering
    \includegraphics[width=\linewidth]{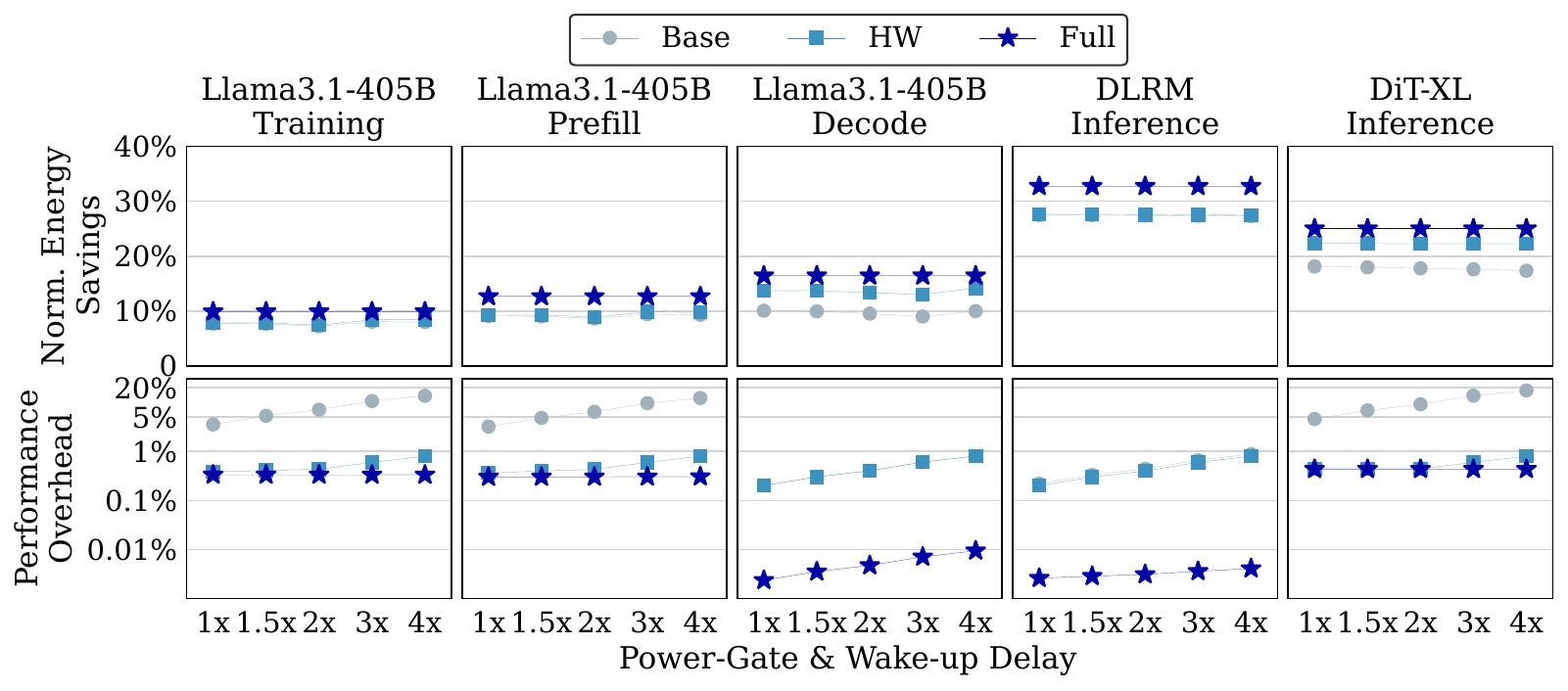}
    \vspace{-4ex}
    \caption{Energy and performance impact of varying power-gate \& wake-up delays. X-axis labels are ratios over the default setting.}
    \label{fig:eval_vary_pg_delay}
\end{figure}

\subsection{Sensitivity Analysis}
\label{sec:eval:sens}

The effect of power gating is affected by various factors, including the threshold voltage (Vth) and the SRAM data retention voltage (DRV), as well as design trade-offs between wake-up delay and leakage reduction.
The voltage parameters are determined by various factors, including feature size, transistor type, and circuit design choices.
For example, for the logic, an aggressive threshold voltage can reduce leakage power in low-power modes, but it may incur longer wake-up delays.
For the SRAM, smaller feature sizes lead to (relatively) higher DRV due to increased sensitivity to manufacturing variations~\mbox{\cite{sram_drv_study}}, making the sleep mode more leaky relative to the OFF mode. Techniques such as assist circuitry have been proposed to reduce leakage, but they come at the cost of additional area or dynamic power~\mbox{\cite{tsmc5nm_sram}}.
This section evaluates \mbox{\pname{}} under different voltage parameters and wake-up delays, as well as different NPU generations with various architectural parameters.

\noindent
\textbf{Varying voltage parameters.}
To study the benefits of \mbox{\pname{}} under different voltage parameters, in \Cref{fig:eval_vary_Vth}, we vary the ratio between the leakage power during OFF/sleep mode and ON mode for the logic and SRAM.
The energy savings decrease as the OFF/sleep mode leakage power becomes relatively larger.
Nevertheless, even under the highest leakage power setting, \full{} outperforms other designs and achieves 4.6\%--16.4\% energy savings.

\noindent
\textbf{Varying power-gate \& wake-up delays.}
\Cref{fig:eval_vary_pg_delay} shows the effect of different power-on/off delays on \pname{}'s energy-saving benefits and performance overheads. In general, a longer delay slightly decreases the energy savings of \pname{} as it increases the break-even time, resulting in fewer power-gating opportunities. It also leads to higher performance overhead in \base{} and \hw{}, but not \full{}, which leverages precise idle period information from the compiler to avoid power-gating a component if the idle period is too short to hide the power-on/off delay.


\begin{figure}
    \centering
    \includegraphics[width=\linewidth]{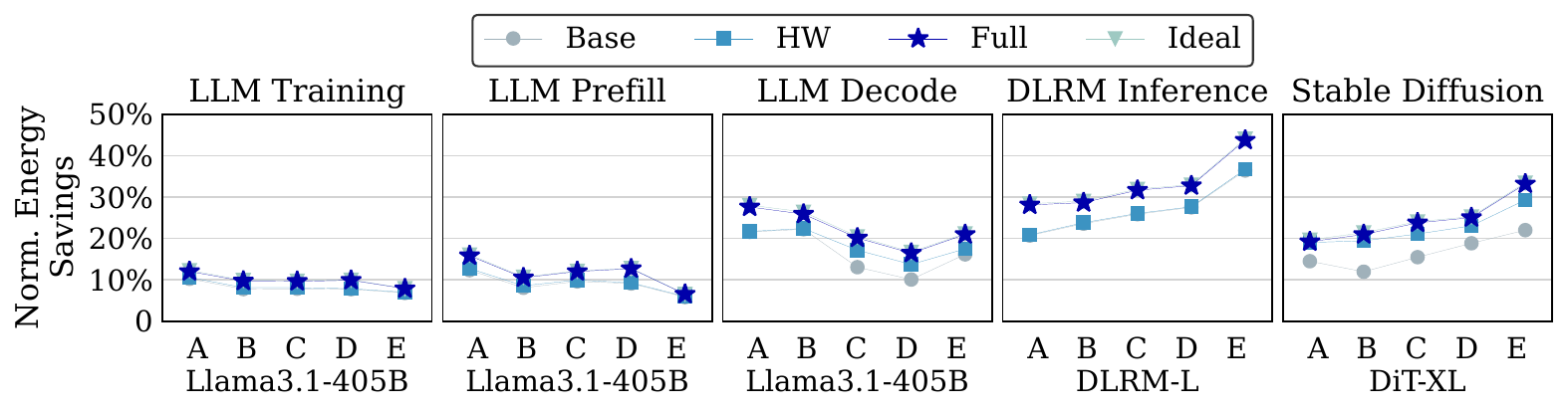}
    \vspace{-4ex}
    \caption{Energy savings of power gating on different NPU generations (normalized to \nopg{}).}
    \label{fig:eval_vary_npu_gen}
\end{figure}

\noindent
\textbf{Varying NPU generations.}
\Cref{fig:eval_vary_npu_gen} compares the benefits of \pname{} on different NPU generations, including a projected future chip, NPU-E (see \Cref{tab:npu_specs}).
\pname{} achieves substantial energy saving on all chips.
Compared to NPU-D, LLM training and prefill, the static energy savings on NPU-E are relatively less since they are compute-intensive and have high dynamic power consumption.
For LLM decode, DLRM, and SD, the energy savings on NPU-E are higher than NPU-D, as NPU-E has larger SRAM (256MB vs. 128MB) and SAs (256$\times$256 vs. 128$\times$128).
The larger units not only consume more static power but also suffer from lower utilization, so the energy savings are more obvious.
Due to the diverse resource demands of ML models, underutilization will remain a problem for future NPUs: ML workloads will still be compute/memory/ICI-bound while underutilizing the non-bottleneck resources.
Power gating alleviates the energy inefficiency due to this fundamental mismatch between workload demand and hardware resources.

\begin{figure}
    \centering
    \includegraphics[width=\linewidth]{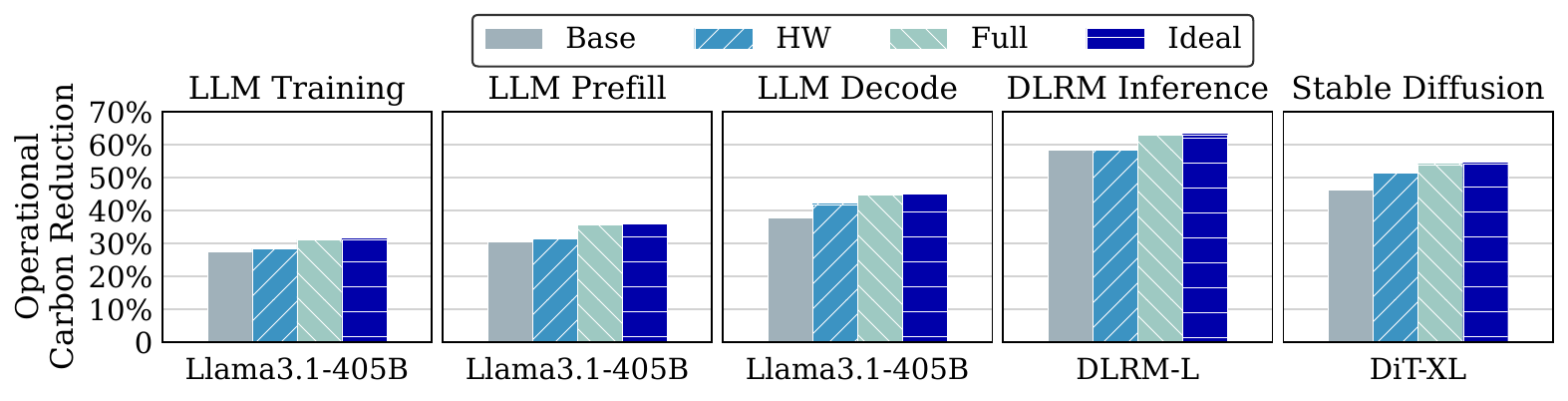}
    \vspace{-4ex}
    \caption{Operation carbon reduction of power gating.}
    \label{fig:eval_carbon}
\end{figure}

\begin{figure*}
    \centering
    \includegraphics[width=\linewidth]{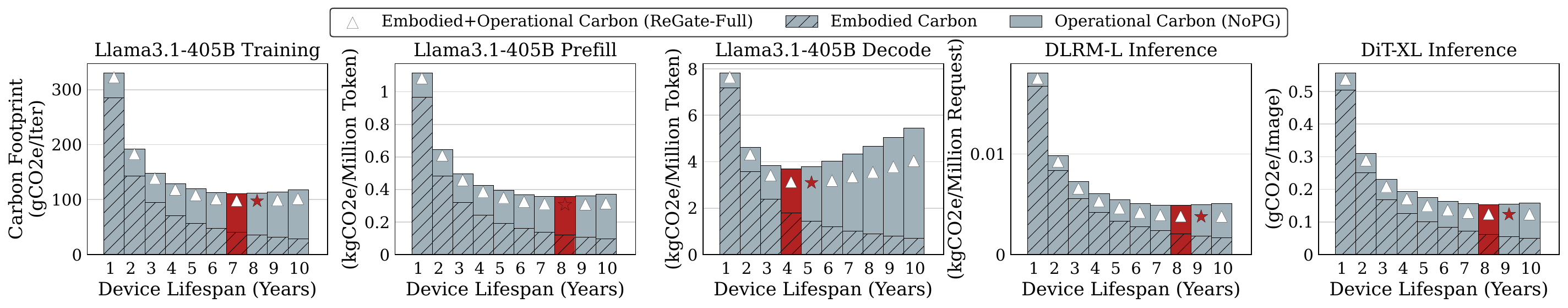}
    \vspace{-4ex}
    \caption{Overall carbon emission of ML workloads over a 10-year period with different device lifespans, assuming the energy efficiency will improve each year by the ratio of NPU-D over NPU-C. We use the embodied carbon for TPUv4/v5p~\cite{carbon_life_cycle}. The optimal lifetime with the lowest carbon is highlighted with red bars and stars. For example, for Llama3.1-405B Decode, the optimal lifespan is 4 years (i.e., we should retire old chips and upgrade to a newer generation every 4 years) without power gating (\nopg{}). With \pname{}, we can extend the lifespan to 5 years.}
    \label{fig:eval_carbon_lca}
\end{figure*}

\subsection{Carbon Efficiency Improvements}
\label{sec:eval:carbon}

The total carbon emission of an NPU chip is the sum of embodied carbon (emission during manufacturing) and operational carbon (emission due to electricity consumption at runtime).
\pname{}'s energy savings directly translate into the reduced operational carbon, as shown in \Cref{fig:eval_carbon}, assuming a carbon intensity of 0.0624 kgCO$_2$e/kWh~\cite{google:env_report:2024} and a 60\% datacenter utilization~\cite{meta:sustainable_ai:mlsys2022}.
With power gating, \pname{} reduces the operational carbon by 31.1\%--62.9\%.
These numbers are much higher than the energy savings in \Cref{fig:eval_energy}, because static power dominates the energy consumption of idle chips, making \pname{}'s benefits more obvious.



We study the impact of embodied carbon in \Cref{fig:eval_carbon_lca}.
If a data center upgrades its NPU fleet too frequently, the embodied carbon will dominate since new NPUs must be acquired frequently.
Reusing old chips will help reduce such costs.
However, since old chips typically have worse energy efficiency than newer chips, their operational carbon emissions will be greater, so there will be no incentive to continue reusing old chips once the operational carbon overhead starts to outweigh the embodied carbon savings.
Exploiting this trade-off leads to an optimal lifespan that achieves the best carbon efficiency, as shown by the red bars and stars in \Cref{fig:eval_carbon_lca}.

The benefits of reducing embodied carbon become more evident when operational carbon represents a smaller portion of total emissions.
Without power gating, the optimal device lifespan is 4--8 years (red bars in \Cref{fig:eval_carbon_lca}). \pname{} extends this range to 5--9 years (red stars in \Cref{fig:eval_carbon_lca}). 
\pname{} demonstrates the effectiveness of power gating in improving the carbon efficiency of NPU chips, paving the way for future sustainable AI chip designs.







\section{Related Work}
\label{sec:related_work}

\noindent
\textbf{Architectural support for power management.}
Power gating techniques have been widely deployed in generic processors. Commercial processors support power-saving modes at core-level (e.g., Intel C-states~\mbox{\cite{intel_cstates}}). 
For caches, prior works employ dynamic cache resizing, power-gating unused cachelines~\cite{gatedvdd:islped00}, or putting cachelines in drowsy mode~\cite{drowsy_cache:isca02}. For execution units, prior works have focused on heuristic- and ML-based idle period detection ~\cite{pg_alu:islped04,itap,charstar} and creating longer idle periods via instruction reordering ~\cite{select_devectorization, gpu_inst_reorder:islped16}. PowerChop~\mbox{\cite{powerchop:isca16}} leverages HW/SW co-design to power-gate non-critical components based on application semantics, it is tailored for the CPU architecture.
There are also prior studies on compiler-directed power optimizations for CPUs~\mbox{\cite{compilerVLIW:isvlsi2016,roy:iscas09}}. They focused on instruction reordering techniques and inserting power management directives based on runtime profiling of component idleness.
\mbox{\pname{}} enables power gating for NPU chips and overcomes the unique architectural obstacles on NPUs. 

Prior works have also investigated power management techniques for AI chips.
Most of them focused on dynamic power, such as dynamic voltage-frequency scaling (DVFS) and clock gating.
They leveraged the power profiles of each DNN layer to select the optimal voltage-frequency setting within the power budget~\mbox{\cite{agrawal:isscc21}} and regulate the peak current using software-controlled pipeline stall signals~\mbox{\cite{kar:isscc24}}.
The Qualcomm Hexagon processor~\mbox{\cite{kalyanam:isscc21}} adjusts the instruction-issue rate of each thread to throttle its power based on thread priorities.
DVFS has also been exploited to adjust the power budget based on input query patterns and latency SLO for ML inference applications~\mbox{\cite{tambe:isscc23}}.
\mbox{\pname{}} focuses on static power and is orthogonal to these techniques.
UPTPU~\mbox{\cite{uptpu}} exploits spatial underutilization for power-gating systolic arrays, but it relies on non-volatile memory (i.e., STT-MRAM) for zero-weight detection and power-gating the weight register, which introduces manufacturing complexity and cost.
\mbox{\pname{}} provides a systematic and practical solution for power-gating all components on an NPU chip.


\noindent
\textbf{System support for power management.}
A few studies to improve the energy efficiency of AI systems have been proposed.
$\mu$-Serve~\cite{micro_serve:atc24}, Envpipe~\cite{envpipe:atc23}, and Perseus~\cite{perseus:sosp24} leverage DVFS to optimize energy consumption for ML training and inference.
Power-aware data center scheduling techniques~\cite{tapas:asplos25,dynamollm:hpca25} route inference requests to evenly distribute cooling and power demands across aisles. \pname{} optimizes the energy consumption of individual NPU chips. It can be deployed in tandem with systems-level solutions for further energy efficiency improvement.

\noindent
\textbf{Carbon-aware AI systems.}
The carbon efficiency of AI is receiving increasing attention in the community.
Prior works analyzed carbon footprints of computer systems~\cite{carbon:hpca21,carbon_life_cycle,carbon_hpc_system}. LLMCarbon~\cite{llm_carbon} and ACT~\cite{carbon_arch_modeling} simulate emissions for different LLMs and AI chips. GreenSKU~\cite{greensku:microsoft:isca24} combines energy-efficient CPUs with reused components to build greener cloud servers. Google investigated the impact of model architecture, training practices, and infrastructure on DNN training emissions~\cite{carbon_large_dnn_training,carbon_plateau_shrink}. EcoServe~\cite{carbon_ecoserve_inference} proposes carbon-aware resource management for LLM serving.
\pname{} contributes by developing architectural solutions to improve the energy efficiency of NPU chips, reducing operational carbon emissions.

\section{Conclusion}
\label{sec:conclusion}
In this paper, we reveal significant power gating opportunities on NPU chips.
We present \pname{}, which enables power gating in NPU chips.
\pname{} supports both HW- and SW-based power gating and enables best-fit decisions via the extended NPU ISA. Our experiments show that
\pname{} improves the energy efficiency of NPUs with negligible performance overhead. 

\begin{acks}
We thank the anonymous reviewers at MICRO'25 for their valuable feedback. We thank Jichuan Chang from Google DeepMind for his insightful discussions throughout this project. 
We thank Yiqi Liu and Michael Wang from the Systems Platform Research Group (Illinois PlatformX) at UIUC for their help with the evaluation and related work investigation.
This work was partially supported by the Hybrid Cloud and AI program at the IBM-Illinois Discovery Accelerator Institute (IIDAI), Google's TPU Research Cloud (TRC) program, and NSF CAREER CNS-2144796. 
\end{acks}

\balance
\bibliographystyle{ACM-Reference-Format}
\bibliography{ref}

\pagebreak
%
%
%
%
%





\appendix
\section{Artifact Appendix}
\label{app:artifact}

\subsection{Abstract}


This artifact contains the NPU simulator used to produce the major results in our paper, including all figures in the power-gating opportunity analysis in $\S$\ref{sec:design_pg:study} and the evaluation of our design in $\S$\ref{sec:eval}. Our artifact provides automated scripts and necessary instructions to reproduce these figures. 
The artifact can be executed on any x86 machine with at least 600 GB of disk space. We strongly recommend running this artifact with at least 128 GB of memory (for a smaller memory, disk swapping may need to be enabled). Optionally, our artifact can run on a Ray cluster with multiple machines to speed up the simulation experiments.

\subsection{Artifact Checklist (Meta-information)}


{\small
\begin{itemize}
  \item {\bf Run-time environment: } Ubuntu 20.04 or newer versions.
  \item {\bf Hardware: } Any x86 machine with at least 600 GB of disk space. At least 48 CPU cores and 128 GB of memory are recommended.
  \item {\bf Metrics: } Energy efficiency, performance, carbon efficiency.
  \item {\bf Output: } Simulator output files and graphs.
  \item {\bf Experiments: } Performance, energy, power, and carbon emission simulation with supplied automation scripts.
  \item {\bf How much disk space required (approximately)?:} 600 GB.
  \item {\bf How much time is needed to prepare workflow (approximately)?:} Less than 10 min.
  \item {\bf How much time is needed to complete experiments (approximately)?:} 160 core-hours.
  \item {\bf Publicly available?:} \url{https://github.com/platformxlab/regate}
  \item {\bf Archived (provide DOI)?:} \url{https://doi.org/10.5281/zenodo.17022414}.
\end{itemize}
}

\subsection{Description}

\subsubsection{How to Access}
The source code can be downloaded from Zenodo at \url{https://doi.org/10.5281/zenodo.17022414}.
For the latest version, we recommend visiting our GitHub repository: \url{https://github.com/platformxlab/regate}.


\subsubsection{Hardware Dependencies}
The artifact can be executed on any x86 machine with at least 600 GB of disk space. We strongly recommend running the artifact on a machine with at least 48 cores and 128 GB of memory (for a smaller memory, disk swapping may need to be enabled). Optionally, the artifact can run on a Ray cluster with multiple machines to speed up the simulation experiments.

\subsubsection{Software dependencies}
The artifact needs a Linux environment (preferably Ubuntu 20.04 or newer).
The artifact uses \texttt{conda} to manage the Python runtime and software dependencies, and uses \texttt{ray} to parallelize the experiments.



\subsection{Installation}
\label{sec:ae:install}

We strongly recommend following the \texttt{README.md} file for detailed instructions on how to install and set up the experiments. We list the steps for installing our artifact and preparing for the experiments as follows:
\begin{enumerate}

    \item (Skip this step if conda is already installed) Install Miniconda: \url{https://www.anaconda.com/docs/getting-started/miniconda/install#linux-terminal-installer}

    \item Clone the repository:
    \begin{lstlisting}[style=BashStyle,escapechar=~~,label=code:instrumentation]
        git clone https://github.com/platformxlab/regate.git
    \end{lstlisting}

    \item Change to the repository directory:
    \begin{lstlisting}[style=BashStyle,escapechar=~~,label=code:instrumentation]
        cd regate
    \end{lstlisting}

    \item Create a conda environment and install the required Python packages:
    \begin{lstlisting}[style=BashStyle,escapechar=~~,label=code:instrumentation]
        conda create --name regate python=3.12.2
        conda activate regate
        pip install -r requirements.txt
    \end{lstlisting}

    \item Under the git repo directory, export the Python path and the simulator root directory:
    \begin{lstlisting}[style=BashStyle,escapechar=~~,label=code:instrumentation]
        export PYTHONPATH=$(pwd):$PYTHONPATH
        export NPUSIM_HOME=$(pwd)
    \end{lstlisting}

    \item Configure the environment variables in \texttt{trace\_util/}\\\texttt{llm\_ops\_generator/run\_scripts/runtime\_env.json}:
    \begin{itemize}
        \item Set \texttt{\$PYTHONPATH} to the root directory of the cloned repository. When running on a single machine, this should be the same as the \texttt{\$PYTHONPATH} set in step 5. 
        \item Set \texttt{RESULTS\_DIR} to \texttt{\$\{NPUSIM\_HOME\}/trace\_util/}\\\texttt{llm\_ops\_generator/results}.
    \end{itemize}

    \item Start ray server:
    \begin{lstlisting}[style=BashStyle,escapechar=~~,label=code:instrumentation]
        ray start --head --port=6379
    \end{lstlisting}

    \item Run the test script to verify the installation:
    \begin{lstlisting}[style=BashStyle,escapechar=~~,label=code:instrumentation]
        cd trace_util/llm_ops_generator/run_scripts
        ./test.sh
    \end{lstlisting}

\end{enumerate}

You may view the progress of the test runs in the Ray dashboard (at http://127.0.0.1:8265/ by default). This may require port forwarding if you are SSH'ing onto a remote machine.

After the script finishes with no errors, under the ``Jobs'' tab in the Ray dashboard, all jobs should have the ``Status'' column showing ``SUCCEEDED''. An output directory will be created at \\\texttt{trace\_util/llm\_ops\_generator/results}.

\subsection{Experiment Workflow}
We strongly recommend following the \texttt{README.md} file for detailed instructions on how to launch the experiments required to reproduce the key figures.
We have provided two automated scripts to reproduce the key figures in the paper.
We strongly recommend running these scripts in a persistent terminal session like \texttt{tmux}, as they may run for hours.
Under the \texttt{trace\_util/llm\_ops\_generator/}\\\texttt{run\_scripts} directory, please run:
\begin{lstlisting}[style=BashStyle,escapechar=~~,label=code:instrumentation]
    ./run_all_ReGate.sh
\end{lstlisting}
and then
\begin{lstlisting}[style=BashStyle,escapechar=~~,label=code:instrumentation]
    ./run_ReGate_sens.sh
\end{lstlisting}
The first script (\texttt{run\_all\_ReGate.sh}) will take approximately 120 core-hours.
It will sweep through all possible NPU pod configurations (NPU version, number of chips, data/tensor/pipeline parallelisms, batch size) for all models, and run the performance, power, and carbon simulation experiments for each configuration. Then, it will run the SLO analysis to pick the best configuration for each model and each NPU version. The output results are used to generate the figures in the power-gating opportunity study ($\S$\ref{sec:design_pg:study}) and evaluation ($\S$\ref{sec:eval}) sections of the paper.

The second script (\texttt{run\_ReGate\_sens.sh}) will take approximately 40 core-hours.
It will run the experiments for the sensitivity analysis figures ($\S$\ref{sec:eval:sens}) in the evaluation section of the paper.

After finishing all experiments, the output results will be stored in the \texttt{trace\_util/llm\_ops\_generator/results} directory by default. Please make sure all jobs finish successfully in the Ray dashboard. If any job fails, please check the logs in the Ray dashboard.

\subsection{Evaluation and Expected Results}
\label{sec:ae:plot_figures}

To reproduce the key figures in the paper, please go to the \\\texttt{trace\_util/llm\_ops\_generator/graphs/graphs\_micro25} directory.
Before plotting the graphs, please make sure to set up the environment variables in step (5) of $\S$\ref{sec:ae:install}.

Then, export the environment variable for the results directory:
\begin{lstlisting}[style=BashStyle,escapechar=~~,label=code:instrumentation]
    export RESULTS_DIR=$NPUSIM_HOME/trace_util/llm_ops_generator/results/
\end{lstlisting}

To generate all figures, simply run \texttt{make}.
Alternatively, \texttt{make motivation} will generate all figures in $\S$\ref{sec:design_pg:study}, and \texttt{make evaluation} will generate all figures in $\S$\ref{sec:eval}.
The generated figures will be under the following directory:
\begin{lstlisting}[style=BashStyle,escapechar=~~,label=code:instrumentation]
    trace_util/llm_ops_generator/graphs/graphs_micro25/outputs
\end{lstlisting}
Please refer to \texttt{trace\_util/llm\_ops\_generator/graphs/}\\\texttt{graphs\_micro25/micro25\_ae\_figures.md} for a checklist of all key figures to be reproduced, along with their corresponding file names. To verify the results, one can compare the generated figures directly with those in the paper.

\subsection{Experiment Customization}

\subsubsection{Output Directory}
To customize the output directory for the simulation results:
\begin{enumerate}
    \item Modify the \texttt{RESULTS\_DIR} variable in \texttt{trace\_util/}\\\texttt{llm\_ops\_generator/run\_scripts/runtime\_env.json}.
    \item In \texttt{trace\_util/llm\_ops\_generator/run\_scripts}, change the export statement for \texttt{RESULTS\_DIR} in all the bash scripts accordingly.
    The \texttt{RESUTS\_DIR} environment variable also needs to be re-exported for plotting the figures in $\S$\ref{sec:ae:plot_figures}.
\end{enumerate}

\subsubsection{Simulation Parameters}
The user can define the NPU hardware and DNN model configurations under the following directory: \texttt{trace\_util/llm\_ops\_generator/configs}.

We provide a set of pre-defined configurations for different DNN models and NPU chips.
The configuration files are used by the \texttt{trace\_util/llm\_ops\_generator/run\_scripts/run\_sim.py} \\ launcher script.
Please see the \texttt{README.md} file for more details.

The power gating parameters are defined in the following file:
\begin{lstlisting}[style=BashStyle,escapechar=~~,label=code:instrumentation]
    trace_util/llm_ops_generator/power_analysis_lib.py
\end{lstlisting}
The user can modify the \texttt{get\_power\_gating\_config()} function to add new power gating configurations, including power gating wake-up/delay cycles and power gating policies for each component.
The \texttt{energy\_operator\_analysis\_main.py} script is invoked for power simulation, and the \texttt{carbon\_analysis\_main.py} script is invoked for carbon emission analysis.
Please refer to the \verb|--help| option of these scripts and the provided automation scripts (\texttt{run\_all\_ReGate.sh} and \texttt{run\_ReGate\_sens.sh}) for how to invoke them with customized simulation parameters.

\subsubsection{Running a Single Tensor Operator}
Please see \texttt{trace\_util/}\\\texttt{llm\_ops\_generator/run\_scripts/run\_single\_op\_main.py} for an example of how to run a single tensor operator simulation.
This tool could be helpful for analyzing a specific operator of interest rather than simulating the entire DNN model.

\subsubsection{Running a Single Experiment}
To run a single experiment, you can directly use the provided ops generator classes.
See the \texttt{trace\_util/llm\_ops\_generator/README.md} for more details.

\subsubsection{Adding New DNN Models}
We currently support LLM, DLRM, DiT, and GLIGEN models. Variants of these models (such as changing the number of layers or hidden dimensions) can be created by adding new configuration files.
To add support for new model architectures, a new model generator class needs to be implemented based on the DNN model's dataflow graph.
Many commonly used operators, such as GEMM, Conv, and LayerNorm, are implemented in \texttt{trace\_util/npusim\_backend/npusim\_lib.py}.
Please refer to the \texttt{trace\_util/llm\_ops\_generator/*\_ops\_generator.py} files as examples on how to implement new model generators.

\subsubsection{Running on a Ray Cluster}
To scale out the simulator on multiple machines, we need to set up a shared storage directory, install the software dependencies on all machines, and configure the Ray cluster. Please refer to the \texttt{README.md} files for more details.


\subsection{Methodology}

Submission, reviewing and badging methodology:

\begin{itemize}
  \item \url{https://www.acm.org/publications/policies/artifact-review-and-badging-current}
  \item \url{https://cTuning.org/ae}
\end{itemize}



\end{document}